\documentclass[fleqn,usenatbib,useAMS]{mnras} 

\usepackage{graphicx}             
\usepackage{amsmath}              
\usepackage{upgreek}              
\usepackage{amssymb}              

\usepackage{bm}                   
\usepackage[mathcal]{euscript}    
\usepackage{CJKutf8}              
\usepackage[T1]{fontenc}          
\usepackage{comment}              
\usepackage{transparent}          
\usepackage{tabularx}             
\usepackage{threeparttable}       
\usepackage{booktabs}             
\usepackage{soul}                 
\usepackage{ulem}                 
\usepackage{multirow}             
\usepackage{makecell}             
\usepackage[dvipsnames]{xcolor}   
\usepackage{tipa}
\usepackage{microtype}            
\DisableLigatures[f]{encoding = *, family = *} 
\usepackage[figuresleft]{rotating}
\emergencystretch=\maxdimen       
\newcommand{\fracp}[2]{\frac{\partial{#1}}{\partial{#2}}}

\newcommand{\hatomega}{\hat{\text{\bf \textscomega}}}

\newcommand{\biborder}[1]{} 

\defcitealias{Baruteau2011}{B11}
\defcitealias{LiYaPing2021}{L21}
\defcitealias{Li2022}{Paper I}

\usepackage{ae,aecompl}
\usepackage{multicol}        

\usepackage{newtxtext,newtxmath}

\title[Binaries Embedded in Discs. II]{\mbox{Hydrodynamical Evolution of Black-Hole Binaries Embedded in AGN Discs:} II. Dependence on Equation of State, Binary Mass,  and Separation Scales}

\author[R. Li and D. Lai]{
  Rixin Li$^{1}$
  \begin{CJK*}{UTF8}{gbsn}
    (李日新)
  \end{CJK*} \thanks{Contact e-mail: \href{mailto:rixin.li@cornell.edu}{rixin.li@cornell.edu}}
  and 
  Dong Lai$^{1}$
  \begin{CJK*}{UTF8}{gbsn}
    (赖东)
  \end{CJK*}
\\
$^{1}$Center for Astrophysics and Planetary Science, Department of Astronomy, Cornell University, Ithaca, NY 14853, USA \\
}

\date{Last updated 2022 XXX XX; in original form 2022 XXX XX}

\pubyear{2022}

\begin{document}
\label{firstpage}
\pagerange{\pageref{firstpage}--\pageref{lastpage}}
\maketitle

\abovedisplayshortskip=6pt plus 3pt minus 2pt 
\belowdisplayshortskip=6pt plus 3pt minus 2pt 
\abovedisplayskip=6pt plus 3pt minus 2pt 
\belowdisplayskip=6pt plus 3pt minus 2pt 

\begin{abstract}
Stellar-mass binary black holes (BBHs) embedded in active galactic nucleus (AGN) discs offer a promising dynamical channel to produce black-hole mergers that are detectable by LIGO/Virgo.  Modeling the interactions between the disc gas and the embedded BBHs is crucial to understand their orbital evolution.  Using a suite of 2D high-resolution simulations of prograde equal-mass circular binaries in local disc models, we systematically study how their hydrodynamical evolution depends on the equation of state (EOS; including the $\gamma$-law and isothermal EOS) and on the binary mass and separation scales (relative to the supermassive BH mass and the Hill radius, respectively).
We find that binaries accrete slower and contract in orbit if the EOS is far from isothermal such that the surrounding gas is diffuse, hot, and turbulent.  The typical orbital decay rate is of the order of a few times the mass doubling rate.
For a fixed EOS, the accretion flows are denser, hotter, and more turbulent around more massive or tighter binaries.
The torque associated with accretion is often comparable to the gravitational torque, so both torques are essential in determining the long-term binary orbital evolution.
We carry out additional simulations with non-accreting binaries and find that their orbital evolution can be stochastic and is sensitive to the gravitational softening length, and the secular orbital evolution can be very different from those of accreting binaries. 
Our results indicate that stellar-mass BBHs may be hardened efficiently under ideal conditions, namely less massive and wider binaries embedded in discs with a non-isothermal EOS.
\end{abstract}

\begin{keywords}
  Compact binary stars(283); Black holes(162); Hydrodynamical simulations(767)
\end{keywords}

\section{Introduction}
\label{sec:intro}

The detection of gravitational waves from the merging black holes (BHs) by the LIGO/Virgo collaboration \citep{Abbott2021arXiv} has motivated many theoretical studies of the formation channels of the BH binaries.  In addition to the isolated binary evolution channel \citep[e.g.,][]{Lipunov1997, Podsiadlowski2003, Belczynski2010, Belczynski2016}, there are several flavors of dynamical channels, including strong gravitational scatterings in dense star clusters \citep[e.g.,][]{PortegiesZwart2000, Miller2009, Samsing2014, Rodriguez2015, Kremer2019}, more gentle ``tertiary-induced mergers'' (often via Lidov-Kozai mechanism) that take place either in stellar triple/quadrupole systems \citep[e.g.,][]{Miller2002, Silsbee2017, LiuLai2018, LiuLai2019, LiuBin2019b, Fragione2019} or in nuclear clusters dominated by a central supermassive BH \citep[e.g.,][]{Antonini2012, Petrovich2017, Hamers2018, LiuBin2019a, LiuLai2020PRD, LiuLai2021}, and (hydro)dynamical interactions in AGN discs \citep[e.g.,][]{Bartos2017, Stone2017, McKernan2018, Tagawa2020a, Samsing2022}.

In the AGN disc scenario, an important question concerns how single BHs can be captured into bound binaries and merge.  \citet{LiLaiRodet2022} showed that when the gas effect is negligible, two BHs in tightly-packed orbits around a supermassive BH become bound to each other in rare, very close encounters due to gravitational wave emission, leading to highly eccentric BH mergers in the LIGO band.  They also showed that weak gas drags do not necessarily enhance the binary capture rate.  Whether strong gas drags can facilitate the capture process is still under active study (Li et al. in prep).

The hydrodynamical evolution of binary black holes (BBHs) in AGN discs is still poorly understood and has only been studied numerically by a handful of previous works.  \citet[][hereafter \citetalias{Baruteau2011}]{Baruteau2011} carried out global simulations in 2D \textit{isothermal} discs and found that a massive (gap-opening) prograde, equal-mass binary is hardened by dynamical friction from the lagging spiral tails trailing each binary component inside the Hill radius.  \citet[][hereafter \citetalias{LiYaPing2021}]{LiYaPing2021} used a similar global setup and found that adequately resolved circum-single disc (CSD) regions in fact lead to expanding binaries.  More works from the same group \citep{LiYaPing2022, Dempsey2022} further found that the BH feedback on CSDs, binary separation, and the vertical structure of CSDs may play an important role in the evolution of the binary.

In \citet[][hereafter \citetalias{Li2022}]{Li2022}, we performed a suite of 2D hydrodynamical simulations of binaries embedded in AGN discs using a co-rotating local disc (``shearing-box'') model.  We targeted realistic BBH to SMBH mass ratios and surveyed a range of binary eccentricities and mass ratios.  We adopted the $\gamma$-law equation of state (EOS) with $\gamma=1.6$ (i.e., far from isothermal).  We found that circular comparable-mass binaries contract, with an orbital decay rate of a few times the mass doubling rate.

When a simplified EOS (such as the $\gamma$-law EOS) is used, the evolution of the binary and the flow dynamics depends only on the dimensionless ratios $R_{\rm H}/H_{\rm g}$ and $R_{\rm H}/a_{\rm b}$ (where $R_{\rm H}$ is the Hill radius, $H_{\rm g}$ is the scale height of the AGN disc far from the binary, and $a_{\rm b}$ is the binary semi-major axis), as well as the binary eccentricity $e_{\rm b}$ and mass ratio $q_{\rm b}$.  In this paper, we use the same numerical scheme as in \citetalias{Li2022} and conduct a survey of parameter space.  We systematically compare the EOS effects on a grid of $R_{\rm H}/H_{\rm g}$ (which is related to the binary mass with respect to the ``thermal mass'' of the disc) and $R_{\rm H}/a_{\rm b}$ values.  We focus on prograde, equal-mass, and circular binaries throughout this paper.  Our goal is to understand how the flow structures around the BBHs and the long-term accretion rate and orbital evolution of BBHs vary in this three dimensional parameter space.

The paper is organized as follows.  In Section \ref{sec:methods}, we recapitulate our numerical scheme and setup.  Section \ref{sec:results} presents the results of our simulations, compares the accretion flow morphologies for various parameters, and describes how the secular binary orbital evolution varies in the parameter space.  We then discuss the relative importance of different torque components on the binary in Section \ref{sec:torques} and evolution of non-accreting binaries in Section \ref{sec:non-accreting}.  Section \ref{sec:summary} summarizes our findings.

\begingroup 
\setlength{\medmuskip}{0mu} 
\setlength\tabcolsep{4pt} 
\setcellgapes{3pt} 
\begin{table*}
  \nomakegapedcells
  \caption{Simulation Setups and Results} \label{tab:runs}
  \makegapedcells 
  \linespread{1.025} 
  \begin{tabular}{lccl|rrrr}
    \hline
    \texttt{Run}
    & $\lambda$
    & $q/h^3$
    & $\gamma$
    & \makecell[c]{$\langle\dot{m}_{\rm b}\rangle$}
    & \makecell[c]{$\langle\dot{\mathcal{E}}_{\rm b}\rangle$}
    & \makecell[c]{$\ell_0$}
    & \makecell[c]{$\displaystyle \frac{\langle\dot{a}_{\rm b}\rangle}{a_{\rm b}}$}
    \\    
    &
    &
    &
    & \makecell[c]{\footnotesize $[\Sigma_{\infty} v_{\rm b} a_{\rm b}]$}
    & \makecell[c]{\footnotesize $\displaystyle \left[\frac{\Sigma_\infty v_{\rm b}^3 a_{\rm b}}{m_{\rm b}}\right]$}
    & \makecell[c]{$[v_{\rm b} a_{\rm b}]$}
    & \makecell[c]{\small $\displaystyle \left[\frac{\langle\dot{m}_{\rm b}\rangle}{m_{\rm b}}\right]$}
    \\
    (1)
    & (2)
    & (3)
    & (4)
    & \makecell[c]{(5)}
    & \makecell[c]{(6)}
    & \makecell[c]{(7)}
    & \makecell[c]{(8)}
    \\
    \hline\hline
    \texttt{II}$^\dagger$
    & $5.0$
    & \makecell[c]{$1$\\$1$\\$1$\\$1$\\$3$\\$3$\\$3$\\$3$\\$9$\\$9$\\$9$\\$9$\\$27$\\$27$\\$27$\\$27$}
    & \makecell[l]{$1.6$\\$1.1$\\$1.001$\\$1.0$\\$1.6$\\$1.1$\\$1.001$\\$1.0$\\$1.6$\\$1.1$\\$1.001$\\$1.0$\\$1.6$\\$1.1$\\$1.001$\\$1.0$}
    & \makecell[r]{$0.17$\\$0.66$\\$0.89$\\$0.90$\\$0.19$\\$1.11$\\$1.65$\\$1.66$\\$0.26$\\$1.94$\\$2.86$\\$2.87$\\$0.29$\\$2.30$\\$3.36$\\$3.33$}
    & \makecell[r]{$-0.48$\\$-0.99$\\$-0.22$\\$-0.18$\\$-0.43$\\$-0.88$\\$0.71$\\$0.64$\\$-0.28$\\$-0.30$\\$1.55$\\$1.47$\\$-0.31$\\$-0.16$\\$0.72$\\$0.61$}
    & \makecell[r]{$-0.23$\\$0.13$\\$0.44$\\$0.45$\\$-0.08$\\$0.30$\\$0.61$\\$0.60$\\$0.23$\\$0.46$\\$0.64$\\$0.63$\\$0.23$\\$0.48$\\$0.55$\\$0.55$}
    & \makecell[r]{$-4.81$\\$-1.99$\\$0.52$\\$0.59$\\$-3.63$\\$-0.59$\\$1.86$\\$1.77$\\$-1.20$\\$0.69$\\$2.09$\\$2.02$\\$-1.13$\\$0.86$\\$1.43$\\$1.37$}
    \\
    \Xhline{3\arrayrulewidth}
    \texttt{III}
    & $7.5$ 
    & \makecell[c]{$1$\\$1$\\$1$\\$1$\\$3$\\$3$\\$3$\\$3$\\$9$\\$9$\\$9$\\$9$\\$27$\\$27$\\$27$\\$27$}
    & \makecell[l]{$1.6$\\$1.1$\\$1.001$\\$1.0$\\$1.6$\\$1.1$\\$1.001$\\$1.0$\\$1.6$\\$1.1$\\$1.001$\\$1.0$\\$1.6$\\$1.1$\\$1.001$\\$1.0$}
    & \makecell[r]{$0.11$\\$0.71$\\$1.02$\\$1.01$\\$0.14$\\$1.41$\\$1.69$\\$1.68$\\$0.24$\\$2.18$\\$3.37$\\$3.36$\\$1.16$\\$2.69$\\$3.92$\\$3.99$}
    & \makecell[r]{$-0.22$\\$-0.63$\\$1.26$\\$1.22$\\$-0.23$\\$0.20$\\$2.03$\\$2.11$\\$-0.13$\\$0.71$\\$3.93$\\$3.74$\\$-1.81$\\$1.58$\\$3.45$\\$3.18$}
    & \makecell[r]{$-0.01$\\$0.28$\\$0.81$\\$0.80$\\$0.10$\\$0.53$\\$0.80$\\$0.81$\\$0.37$\\$0.58$\\$0.79$\\$0.78$\\$0.11$\\$0.65$\\$0.72$\\$0.70$}
    & \makecell[r]{$-3.09$\\$-0.77$\\$3.46$\\$3.40$\\$-2.21$\\$1.28$\\$3.40$\\$3.52$\\$-0.08$\\$1.65$\\$3.33$\\$3.22$\\$-2.12$\\$2.18$\\$2.76$\\$2.59$}
    \\
    \Xhline{3\arrayrulewidth}
    \texttt{IV}
    & $10$ 
    & \makecell[c]{$1$\\$1$\\$1$\\$1$\\$3$\\$3$\\$3$\\$3$}
    & \makecell[l]{$1.6$\\$1.1$\\$1.001$\\$1.0$\\$1.6$\\$1.1$\\$1.001$\\$1.0$}
    & \makecell[r]{$0.08$\\$0.89$\\$0.90$\\$0.90$\\$0.14$\\$1.33$\\$1.45$\\$1.43$}
    & \makecell[r]{$-0.14$\\$-0.11$\\$0.82$\\$0.81$\\$-0.18$\\$0.88$\\$2.18$\\$2.22$}
    & \makecell[r]{$0.07$\\$0.47$\\$0.73$\\$0.72$\\$0.18$\\$0.66$\\$0.87$\\$0.89$}
    & \makecell[r]{$-2.47$\\$0.76$\\$2.82$\\$2.79$\\$-1.58$\\$2.32$\\$4.00$\\$4.10$}
    \\
    \Xhline{3\arrayrulewidth}
    \texttt{V}
    & $12.5$ 
    & \makecell[c]{$1$\\$1$\\$1$\\$1$\\$3$\\$3$\\$3$\\$3$}
    & \makecell[l]{$1.6$\\$1.1$\\$1.001$\\$1.0$\\$1.6$\\$1.1$\\$1.001$\\$1.0$}
    & \makecell[r]{$0.07$\\$0.84$\\$0.87$\\$0.88$\\$0.10$\\$1.24$\\$1.69$\\$1.66$}
    & \makecell[r]{$-0.15$\\$-1.71$\\$-1.93$\\$-2.00$\\$-0.27$\\$-2.74$\\$-3.97$\\$-3.81$}
    & \makecell[r]{$-0.01$\\$-0.01$\\$-0.05$\\$-0.07$\\$-0.14$\\$-0.05$\\$-0.09$\\$-0.07$}
    & \makecell[r]{$-3.11$\\$-3.07$\\$-3.42$\\$-3.53$\\$-4.11$\\$-3.42$\\$-3.71$\\$-3.59$}
    \\
    \hline
  \end{tabular} \\
  \begin{flushleft}
    {\large N}OTE ---Columns: 
    (1) run names; 
    (2) ratio between binary semi-major axis $a_{\rm b}$ and $R_{\rm H}\equiv R q^{-1/3}$, i.e., $\lambda = R_{\rm H}/a_{\rm b}$;
    (3) $q/h^3 = (m_{\rm b}/M)(H_{\rm g}/R)^-3 = (R_{\rm H}/H_{\rm g})^3$;
    (4) $\gamma$ in the $\gamma$-law EOS (the cases with $\gamma=1$ are isothermal runs);  
    (5) time-averaged accretion rate;
    (6) time-averaged rate of change in binary specific energy;
    (7) accretion eigenvalue;
    (8) binary semimajor axis change rate or migration rate.
    \\[0.25em]
    $^\dagger$ --- We reserve the name \texttt{Run I} for $\lambda=2.5$ so the run names are consistent with those in \citetalias{Li2022}. 
    \end{flushleft}
\end{table*}
\endgroup

\section{Methods}
\label{sec:methods}

To study the hydrodynamical evolution of binaries embedded in accretion discs (see the schematic cartoon in Fig. 1 in \citetalias{Li2022}), we use the code \texttt{ATHENA} \citep{Stone2008, Stone2010} with a similar setup to \citetalias{Li2022}.  Section \ref{subsec:schemes} briefly reiterates our numerical model.  Section \ref{subsec:setups} summarizes the parameter choices for our simulations.

\begin{figure*}
  \centering
  \includegraphics[width=\linewidth]{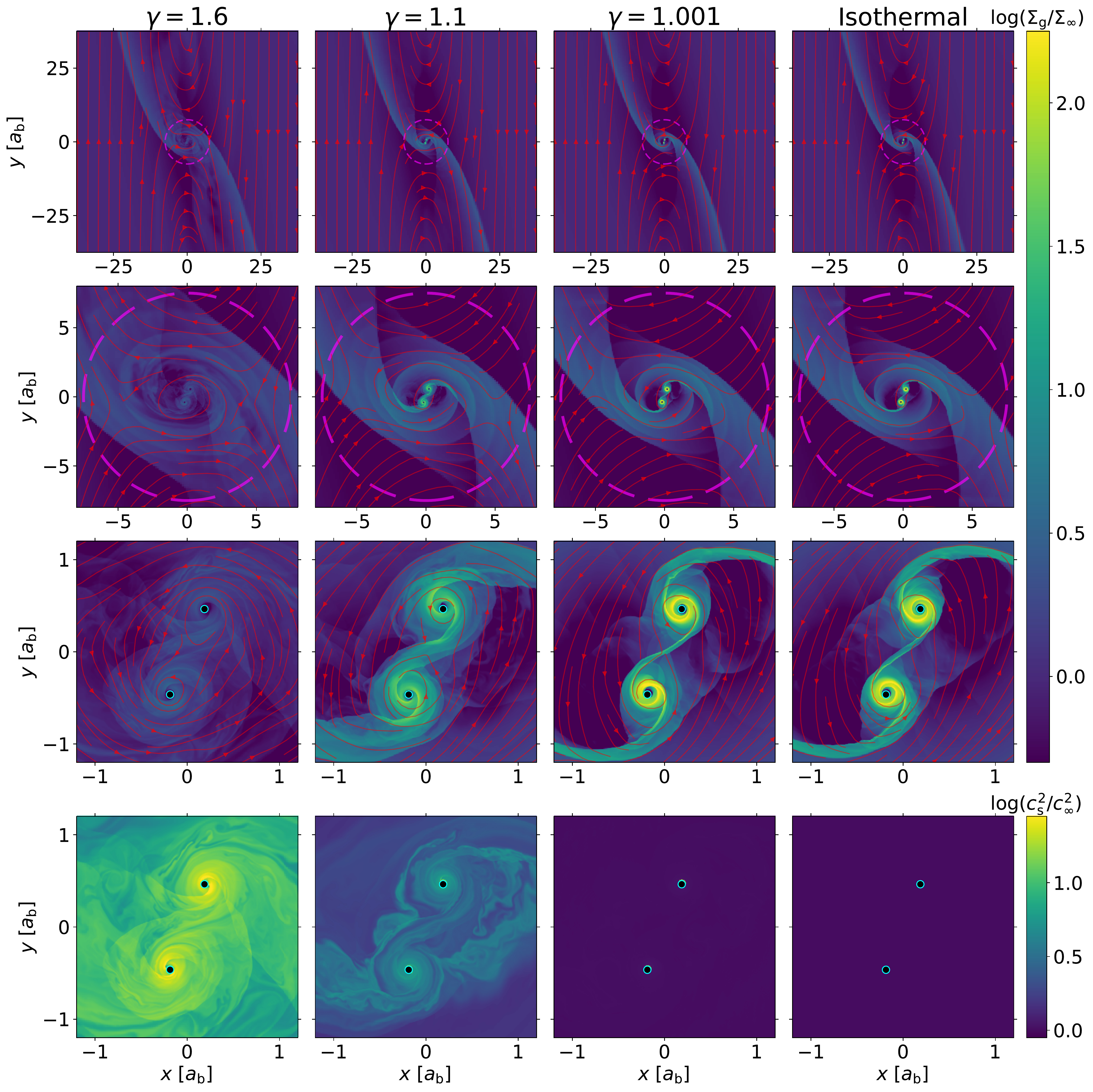}
  \caption{Final snapshots for \texttt{Run III} ($\lambda = 7.5$) with $q/h^3=3$ and (from \textit{left} to \textit{right}) $\gamma=1.6$, $1.1$, $1.001$, and isothermal, where the mesh is refined progressively towards the binary (zooming in from \textit{top} to \textit{bottom}), showing the gas surface density and detailed flow structures (\textit{red} streamlines) in the first three rows and the thermal structures in the bottom row.  The \textit{magenta dashed} circles in the first two rows with a radius of $R_{\rm H}$ denotes the Hill radius of the binary.  The \textit{cyan solid} circles in the last two rows with a radius of $r_{\rm s} = 0.04 a_{\rm b}$ represent the sink radius of each binary component.  \label{fig:snap_runIII_qth3}}
\end{figure*}

\begin{figure*}
  \centering
  \includegraphics[width=\linewidth]{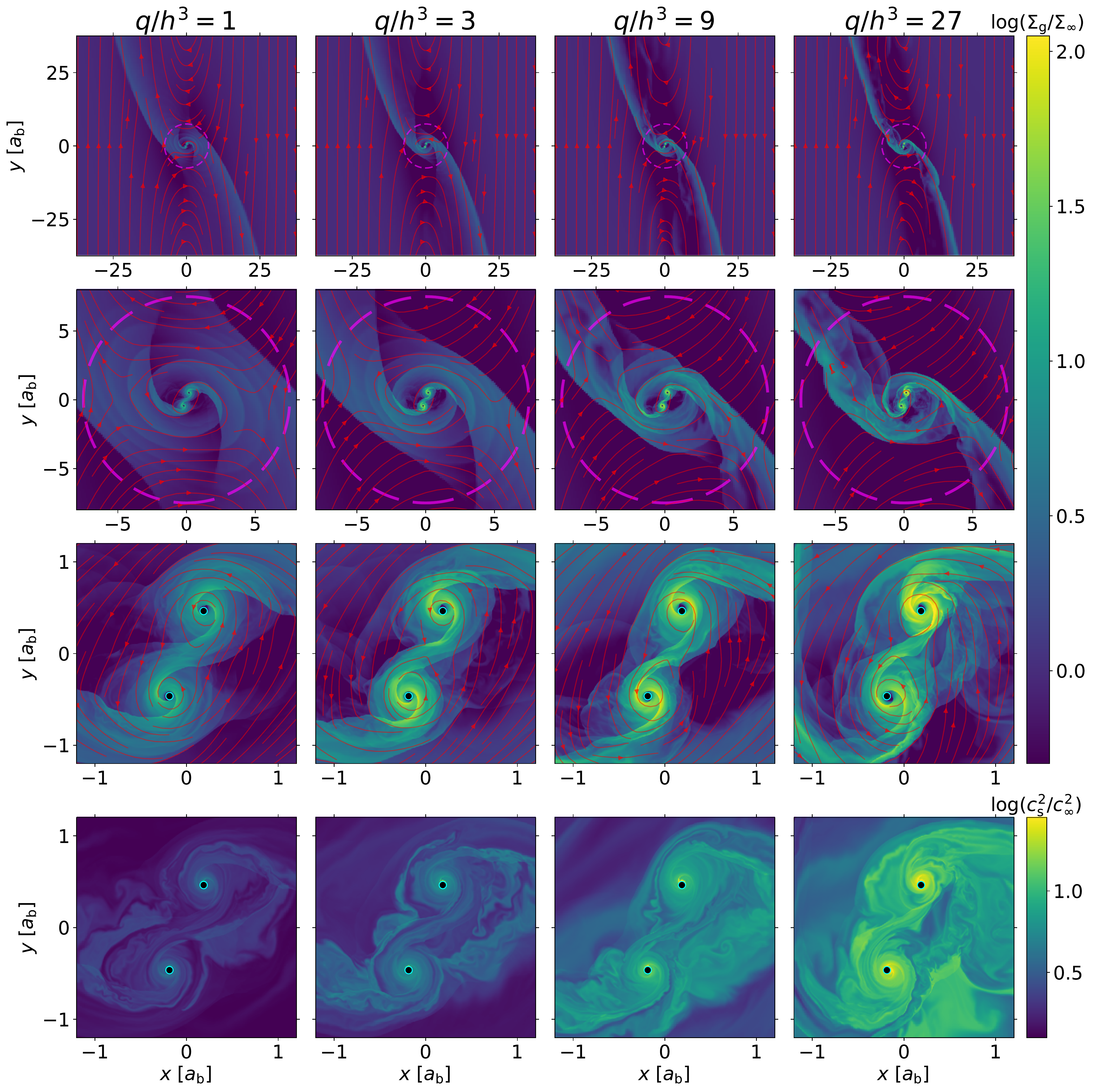}
  \caption{Similar to Fig. \ref{fig:snap_runIII_qth3} but for \texttt{Run III} ($\lambda = 7.5$) with $\gamma=1.1$ and (from \textit{left} to \textit{right}) $q/h^3=1$, $3$, $9$, and $27$.  \label{fig:snap_runIII_ga1.1}}
\end{figure*}

\begin{figure*}
  \centering
  \includegraphics[width=\linewidth]{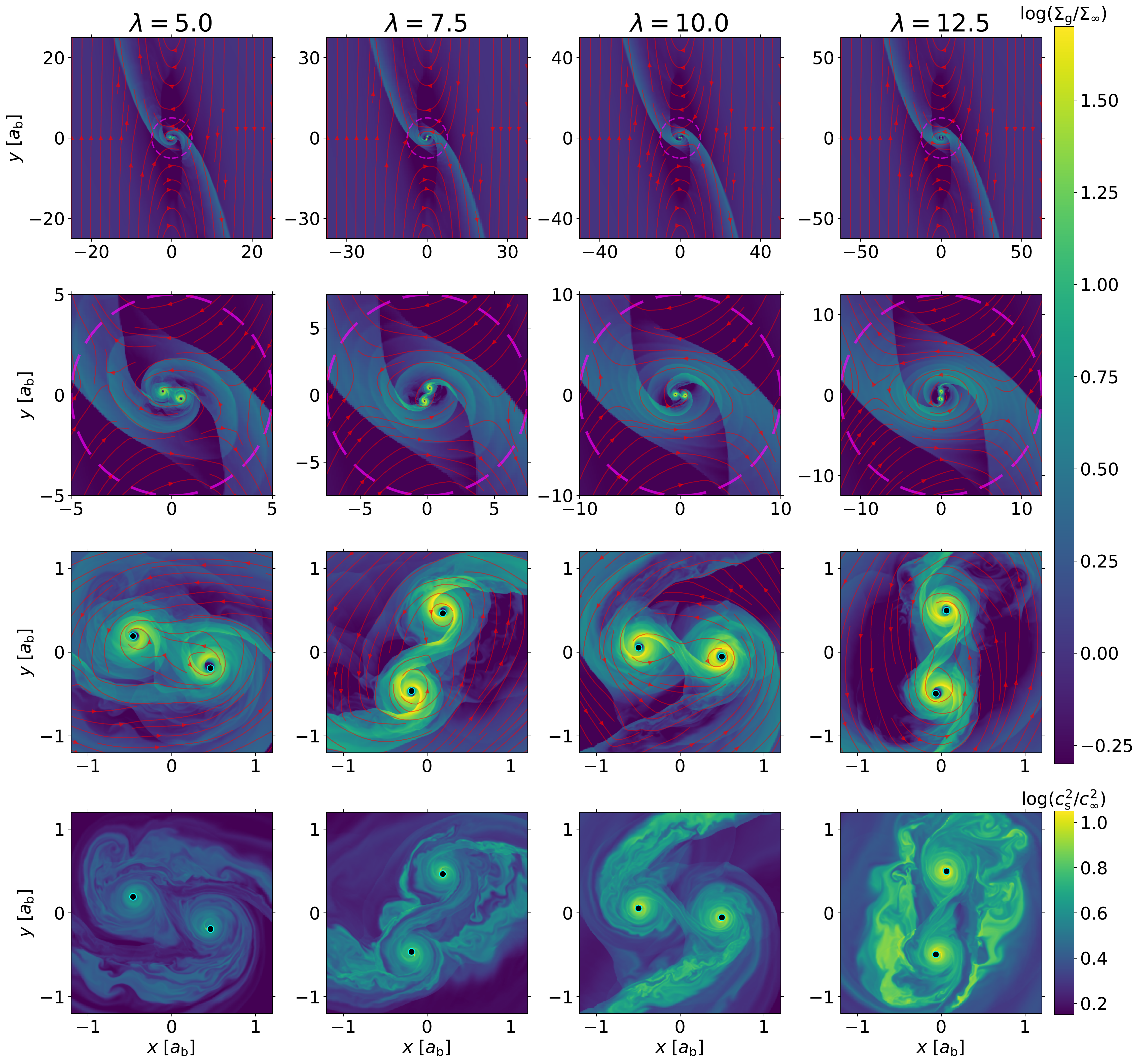}
  \caption{Similar to Fig. \ref{fig:snap_runIII_qth3} but for cases with $q/h^3=3$, $\gamma=1.1$, and (from \textit{left} to \textit{right}) $\lambda=5$, $7.5$, $10$, and $12.5$.  Note that the plotting range varies by column in the first two rows to accommodate different domain sizes and $R_{\rm H}$.  \label{fig:snap_qth3_ga1.1}}
\end{figure*}

\subsection{Simulation Setup}
\label{subsec:schemes}

We consider a binary (with component masses $m_1$ and $m_2$) centred in a small patch of an accretion disc around a massive object (e.g., a super massive black hole (SMBH) with mass $M$) using the local shearing box approximation \citep{Goldreich1965, Hawley1995, Stone2010}.  The centre of mass (COM) of the binary, i.e., the centre of the computational domain --- $(x, y) = (0, 0)$ --- is located at a fiducial disc radius $R$ from the SMBH.  At this location, the Keplerian velocity is $V_{\rm K}=\sqrt{GM/R}$ and the Keplerian frequency is $\Omega_{\rm K}=V_{\rm K}/R$.  Our reference frame rotates at this frequency.

In the rotating frame, we simulate the dynamics of an inviscid compressible flow by solving the following equations of gas dynamics in 2D
\begin{align}
  \fracp{\Sigma_{\rm g}}{t} + \nabla \cdot \left(\Sigma_{\rm g} \bm{u}\right) &= 0, \label{eq:gascon} \\
  \begin{split}\label{eq:gasmom}
    \fracp{(\Sigma_{\rm g} \bm{u})}{t} + \nabla\cdot(\Sigma_{\rm g} \bm{u}\bm{u} + P\bm{I}) &=\\
    \Sigma_{\rm g} \biggl[ 2\bm{u}\times\bm{\Omega}_{\rm K} &+ 2 q_{\rm sh} {\Omega}_{\rm K}^2 \bm{x} - \nabla \phi_{\rm b} \biggr], 
  \end{split} \\
  \begin{split}\label{eq:gasE}
    \fracp{E}{t} + \nabla\cdot\left[ (E + P) \bm{u}\right] &= \Sigma_{\rm g} \bm{u} \cdot \left(2 q_{\rm sh}\Omega_{\rm K}^2 \bm{x} - \nabla\phi_{\rm b} \right) , \\
    E = \frac{P}{\gamma - 1} &+ \frac{1}{2} \Sigma_{\rm g} (\bm{u} \cdot \bm{u}),
  \end{split}
\end{align}
where $\Sigma_{\rm g}$, $\bm{u}$, $P$, $E$, and $\gamma$ are surface density, velocity, pressure, total energy surface density, and adiabatic index of gas, $\bm{I}$ is the identity matrix, $\bm{\Omega}_{\rm K}$ aligns with $\hat{\bm{z}}$, $q_{\rm sh} \equiv \mathbf{-} \textnormal{d}\ln \Omega_{\rm K}/\textnormal{d} \ln R$ is the background shear parameter and is $3/2$ for a Keplerian disc, $\phi_{\rm b}$ is the gravitational potential of the binary
\begin{equation}
    \phi_{\rm b}(\bm{r}_{k}) = -\frac{G m_1}{|\bm{r}_1 - \bm{r}_k| + \xi_{\rm s}} -\frac{G m_2}{|\bm{r}_2 - \bm{r}_k| + \xi_{\rm s}},
\end{equation}
where $\bm{r}_1$ and $\bm{r}_2$ denote the position vectors of the binary components, $\bm{r}_k$ is the centre position of the $k$-th cell in the computational domain, and $\xi_{\rm s}$ is the gravitational softening length.

In this work, we consider both the $\gamma$-law EOS and the isothermal EOS.  For the $\gamma$-law EOS, the sound speed of the gas is given by $c_{\rm s} = \sqrt{\gamma P / \Sigma_{\rm g}}$.  For the isothermal EOS, we do not evolve the energy equation (i.e., Eq. \ref{eq:gasE}) and adopt $P = c_{\rm s}^2 \Sigma_{\rm g}$.  Below we sometimes use $\gamma=1$ to indicate the isothermal EOS for conciseness. 

The binary in our models has total mass $m_{\rm b} = m_1 + m_2$ and orbits on a prescribed circular orbit with a semi-major axis of $a_{\rm b}$.  The mean orbital frequency, orbital angular momentum, and energy in the inertial frame are thus
\begin{align}
  \bm{\Omega}_{\rm b} &= \sqrt{\frac{G m_{\rm b}}{a_{\rm b}^3}}\ \hatomega_{\rm b} = \frac{v_{\rm b}}{a_{\rm b}} \hatomega_{\rm b}, \qquad \text{with}~ v_{\rm b} \equiv \sqrt{\frac{G m_{\rm b}}{a_{\rm b}}}, \\
  \bm{L}_{\rm b} &= \mu_{\rm b} \bm{\ell}_{\rm b} = \mu_{\rm b} \bm{\Omega}_{\rm b} a_{\rm b}^2, \\
  E_{\rm b} &= \mu_{\rm b} \mathcal{E}_{\rm b} = -\mu_{\rm b} \frac{G m_{\rm b}}{2 a_{\rm b}},
\end{align}
where $\hatomega_{\rm b}$ is binary normal unit vector, $\mu_{\rm b} = m_1 m_2 / m_{\rm b}$ is the reduced mass, and $\ell_{\rm b}$ and $\mathcal{E}_{\rm b}$ are the specific angular momentum and specific energy, respectively.  Throughout this paper, we consider co-planar, prograde, and equal-mass binaries on prescribed orbits (see Section 2.1 of \citetalias{Li2022} for detailed prescriptions).

The binary interacts with the background flow, which, in the shearing box frame, is given by
\begin{equation}\label{eq:v_wind}
  \begin{aligned}
    \bm{V}_{\rm w} &= \bm{V}_{\rm sh} + \bm{\Delta V}_{\rm K} \\
    &= -\frac{3}{2} \Omega_{\rm K} x \bm{\hat{y}} - \beta h^2 V_{\rm K} \bm{\hat{y}}
  \end{aligned}
\end{equation}
far from the binary, where $\bm{V}_{\rm sh}(x)$ denotes the Keplerian shear, $\bm{\Delta V}_{\rm K}$ is the deviation from Keplerian velocity, $h\equiv H_{\rm g}/R$ is the disc aspect ratio (with $H_{\rm g}$ the disc scale height), and $\beta \simeq - \textnormal{d}\ln P / \textnormal{d}\ln R $ is an order unity coefficient determined by the (background) disc pressure profile.

There are several length scales that are relevant to our problem:
\begin{itemize}
  \item Binary semi-major axis $a_{\rm b}$;
  \item Hill radius $R_{\rm H} \equiv R (m_{\rm b}/M)^{1/3} \equiv R q^{1/3}$ (note that this definition of Hill radius differs from the standard one $R_{\rm H}' = R_{\rm H}/3^{1/3}$);
  \item Bondi radius $R_{\rm B} = G m_{\rm b} / c_{\rm s,\infty}^2$, where $c_{\rm s,\infty}$ is the sound speed of the background gas (far from the binary);
  \item Scale height of the background disc $H_{\rm g} = c_{\rm s,\infty} / \Omega_{\rm K}$.
\end{itemize}
However, their ratios depend on only two dimensionless parameters:
\begin{align}
  \frac{R_{\rm H}}{H_{\rm g}} &= \left( \frac{R_{\rm B}}{H_{\rm g}} \right)^{1/3} = \left( \frac{q}{h^3} \right)^{1/3}, \label{eq:R_H_over_H} \\
  \frac{R_{\rm H}}{a_{\rm b}} &= \lambda. \label{eq:R_H_over_a_b}
\end{align}
Similarly, the relevant velocity scales are $c_{\rm s,\infty}$, $v_{\rm b}$, the velocity shear across the binary $V_{\rm s}\equiv |V_{\rm sh}(x=a_{\rm b})| = \frac{3}{2}\Omega_{\rm K} a_{\rm b}$, and $|\Delta V_{\rm K}|$.  The first three are related by the same dimensionless parameters:
\begin{align}
  \frac{v_{\rm b}}{c_{\rm s,\infty}} &= \lambda^{1/2} \left(\frac{q}{h^3} \right)^{1/3}, \\
  \frac{V_{\rm s}}{c_{\rm s,\infty}} &= \frac{3}{2\lambda} \left(\frac{q}{h^3} \right)^{1/3}.
\end{align}
For thin discs ($h \ll 1$), $|\Delta V_{\rm K}|$ is very subsonic ($|\Delta V_{\rm K}|/c_{\rm s,\infty} = |\beta|h \ll 1$) and is typically much smaller than $V_{\rm s}$
\footnote{In our calculations, we set $h=0.01$, which guarantees that $|\Delta V_{\rm K}|\ll V_{\rm s}$ for all $q/h^3$ and $\lambda$ values surveyed in this paper.  If the disc has a large thickness, e.g., $h=0.05$, then $|\bm{\Delta V}_{\rm K}|$ may become comparable to $\bm{V}_{\rm s}$, leading to profoundly asymmetric shear flows in the vicinity of the binary.}

Therefore, we expect our results, when appropriately scaled, depend on various physical quantities only through two dimensionless parameters: $q/h^3$ and $\lambda$.  The former parameter, $q/h^3$, is also known as the ratio of the binary mass $m_{\rm b}$ to the so-called thermal mass, $h^3 M$.

To evaluate the accretion onto the binary components and to calculate the torque, energy transfer rate and orbital evolution of the binary, we treat each binary component as an absorbing sphere with a sink radius of $r_{\rm s}$ and calculate all quantities along an evaluation radius $r_{\rm e}$ immediately outside $r_{\rm s}$ (see Section 2.2 of \citetalias{Li2022} for further details; see also Section \ref{sec:non-accreting} for our simulation results with non-accreting binaries).

\subsection{Numerical Parameters}
\label{subsec:setups}

The flow dynamics and results of our simulations depend on the dimensionless parameters ($q/h^3$, $\lambda$; see Eqs. \ref{eq:R_H_over_H} and \ref{eq:R_H_over_a_b}) and the EOS ($\gamma$).  Tables \ref{tab:runs} summarizes the parameters for all of our runs.  For $\lambda=5$ and $7.5$ (\texttt{Run II} and \texttt{Run III} series, respectively), we survey a range of $\gamma$ ($1.6$, $1.1$, $1.001$, and isothermal) and a range of $q/h^3$ ($1$, $3$, $9$, and $27$).  For $q/h^3=1$ and $3$, we further survey a range of $\lambda$ ($5$, $7.5$, $10$, $12.5$; \texttt{Run II}-\texttt{IV} series, respectively) with the same range of $\gamma$ ($1.6$, $1.1$, $1.001$, and isothermal).

Similar to \citetalias{Li2022}, the gas in all of our simulations are initialized with $\Sigma_{\rm g} = \Sigma_{\infty}$ and with the velocity given by the background wind profile $\bm{V}_{\rm w}$ (see Eq. \ref{eq:v_wind}).  We set the root computational domain size to $10 R_{\rm H}$ in both $x$ and $y$ directions.  At all boundaries of the root domain, we adopt a wave-damping open boundary condition (BC) that damps the flow back to its initial state with a wave-damping timescale $P_{\rm d} = 0.02\Omega_{\rm b}^{-1}$.  To properly resolve the flow around the binary, we employ multiple levels of static mesh refinement (SMR) towards the binary, where the resolution at the finest level is $a_{\rm b}/\delta_{\rm fl} = 245.76$, where $\delta_{\rm fl}$ is the cell size at that level.  In all simulations, we adopt a sink radius of $r_{\rm s}=0.04 a_{\rm b}$, an evaluation radius of $r_{\rm e}=r_{\rm s}+\sqrt{2} \delta_{\rm fl}$, and a gravitational softening length of $10^{-8} a_{\rm b}$.

In each simulation, we prescribe the binary orbital motion and evolve the flow dynamics for $500\Omega_{\rm b}^{-1}$ for the cases with $\lambda=5$, for $1000\Omega_{\rm b}^{-1}$ for the cases with $\lambda=7.5$, and for $2000\Omega_{\rm b}^{-1}$ for the cases with $\lambda=10$ and $12.5$.
With the accretion rates and torques measured on-the-fly in each time-step, the long-term binary orbital evolution is determined by the time-averaged long-term measurements in the post-processing analyses (see Section 2.2 in \citetalias{Li2022}).  The time-averaging is done over the last $240\Omega_{\rm b}^{-1}$ / $480\Omega_{\rm b}^{-1}$ / $960\Omega_{\rm b}^{-1}$ for runs that evolve $500\Omega_{\rm b}^{-1}$ / $1000\Omega_{\rm b}^{-1}$ / $2000\Omega_{\rm b}^{-1}$.
For the particular case with $(\lambda, q/h^3, \gamma)=(7.5, 27, 1.6)$, the accretion flows around the binary are much more violent since $v_{\rm b}$ is highly supersonic and the CSDs heat up easily.  To accommodate this situation, we evolve it for $2000\Omega_{\rm b}^{-1}$ such that a longer time window is available for analysis.

\section{Results}
\label{sec:results}

Tables \ref{tab:runs} summarizes the key parameters and results of our simulation suite in a three dimensional parameter space, namely $\lambda$, $q/h^3$, and $\gamma$.  We first describe the accretion flow morphologies in Section \ref{subsec:flow_field} and then present the orbital evolution results in Section \ref{subsec:orbital_evolution}.

\subsection{Flow Structure}
\label{subsec:flow_field}

\citetalias{Li2022} has demonstrated that the quasi-steady state flow structures outside $R_{\rm H}$ barely change due to binary configuration (e.g., $\lambda$, binary eccentricity and internal mass ratio).  Fig. \ref{fig:snap_runIII_qth3} shows that this finding remains valid for different choices of the EOS, where that the grand half bow shocks have broadly similar morphologies far away from the binary.

For flows inside $R_{\rm H}$, we find that the CSDs become more massive, more compact, and cooler as $\gamma$ decreases.  When $\gamma=1.6$, the CSDs are hot and puffed up, pushing the circumbinary flows outward.  On the contrary, the CSDs in the isothermal cases and the $\gamma=1.001$ cases are effectively cooled and are able to maintain gas densities that are orders of magnitude higher.  The fact that the $\gamma=1.001$ cases exhibit good agreement with the isothermal cases on flow morphologies reassures us the numerical robustness.  In addition, we notice that the small spiral shocks in the CSDs becomes wider as $\gamma$ increases.  In other words, the spirals wind more tightly with a smaller pitch angle in cases with a smaller $\gamma$.

Fig. \ref{fig:snap_runIII_ga1.1} shows that the circumbinary flows and the CSDs are more turbulent, denser, and hotter as $q/h^3$ increases.  The Mach number of the orbital velocity $v_{\rm b}/c_{\rm s, \infty}$ increases from $2.74$ to $8.22$ when $q/h^3$ increases from $1$ to $27$ in \texttt{Run III} series ($\lambda=7.5$), leading to the transition from relatively laminar flows to turbulent flows.  Moreover, a higher $q/h^3$ represents a deeper potential well around each binary component, which naturally results in a more massive CSD, where more potential energy dissipates into heat.  The consequence of the high Mach numbers in fact extends beyond $R_{\rm H}$, where the cones of the grand half bow shocks become slightly narrower, as expected for Bondi-Hoyle-Lyttleton-like accretion with higher velocities.

Figs. \ref{fig:snap_qth3_ga1.1} explores an extended range of $\lambda$ and once more confirms that the binary separation has little influence on the flow outside $R_{\rm H}$.  Comparing to $R_{\rm H}$, the binary separation by definition becomes smaller as $\lambda$ increases, where the circumbinary flow structures resemble more closely with the typical flow pattern around a single object \citep{Tanigawa2002}, i.e., a pair of spiral shock winding from the binary towards $R_{\rm H}$ and forming a pair of shock triple points when joining with the pair of grand half bow shocks.  For larger $\lambda$, the spiral shocks from the binary as a whole
\footnote{In \citetalias{Li2022}, we referred to such large spiral shocks as ``a small half bow shock''.  These two terms are used interchangeably in this paper.}
are more steady due to the less intertwinement with the compact CSDs and their small spiral shocks originated from each binary component.

Furthermore, the CSDs becomes slightly more massive and hotter as $\lambda$ increases.  Because a smaller binary separation moderately deepens the potential well around each component such that its CSD is able to sustain more gas mass.  Moreover, a closer binary has a higher orbital velocity.  The corresponding Mach number increases from $3.22$ to $5.10$ as $\lambda$ increases from $5$ to $12.5$ (with $q/h^3=3$), again leading to more turbulent CSDs with more heat.  We also find that the fast moving binaries ($\lambda \gtrsim 10$) are able to leave voids (i.e., regions with very low surface density but high temperature) behind along their orbit, whereas such regions are quickly refilled and then cooled by the ambient gas around wider binaries (i.e., smaller $\lambda$).

Since the accretion flow morphology strongly depends on $\gamma$, $q/h^3$, and $\lambda$, similarly strong dependences of accretion rate and binary orbital evolution on these parameters are expected, which we address in the next section.

\begin{figure*}
  \centering
  \includegraphics[width=0.495\linewidth]{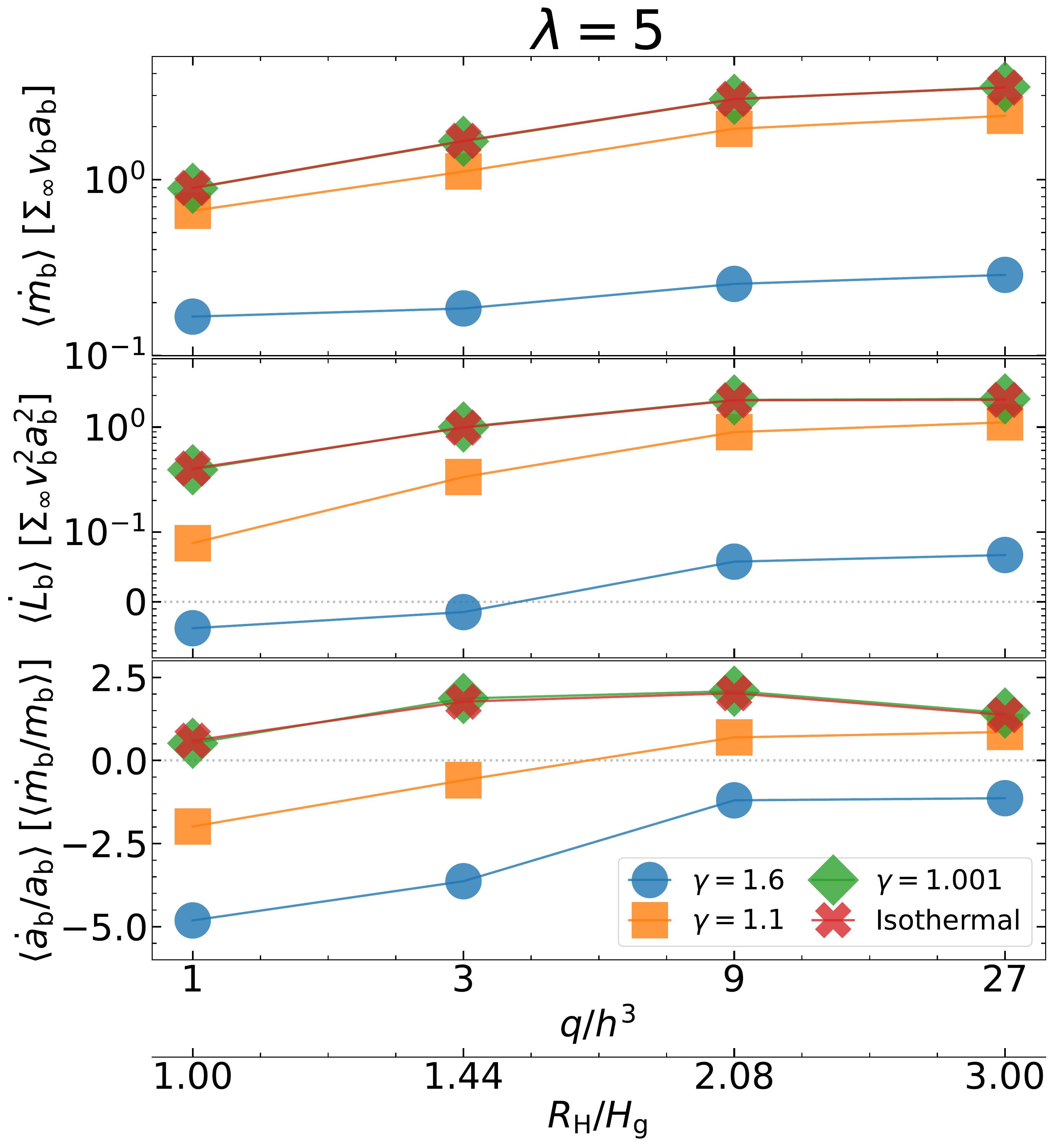}
  \includegraphics[width=0.495\linewidth]{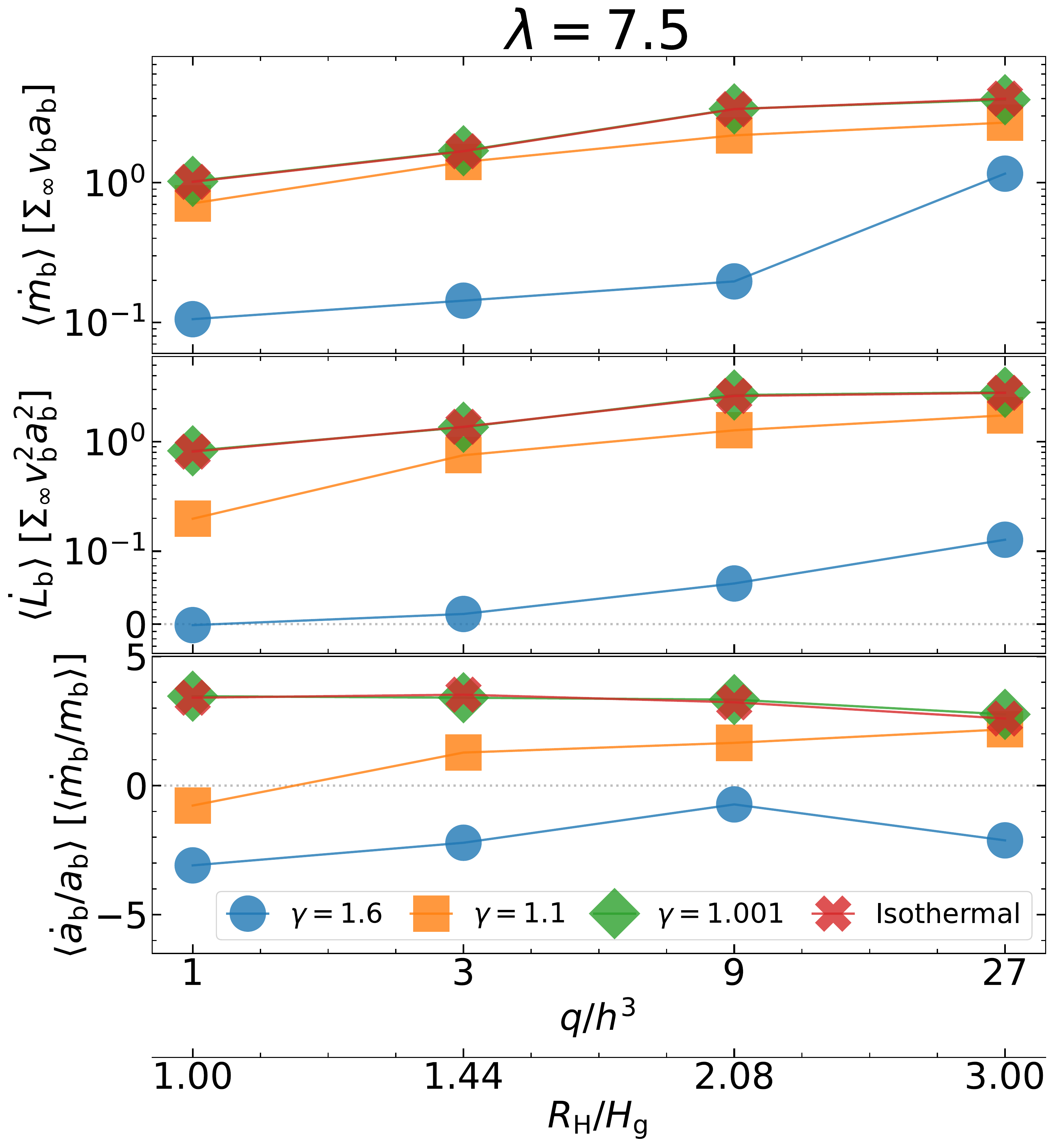}
  \caption{Time-averaged measurements of (from \textit{top} to \textit{bottom}) accretion rate $\langle\dot{m}_{\rm b}\rangle$, total torque $\langle\dot{L}_{\rm b}\rangle$, and binary migration rate $\langle\dot{a}_{\rm b}\rangle/a_{\rm b}$ as a function of $q/h^3$, color-coded by the EOS ($\gamma=1.6$: \textit{blue circle}; $1.1$: \textit{orange square}; $1.001$: \textit{green diamond}; isothermal: \textit{red cross}), for \texttt{Run II} series ($\lambda=5$, \textit{left}) and \texttt{Run III} series ($\lambda=7.5$, \textit{right}).  In the middle panels for $\langle\dot{L}_{\rm b}\rangle$, the $y$-axis combines linear (\textit{below} $10^{-1}$) and logarithmic (\textit{above} $10^{-1}$) scales to better show the trends in $\gamma=1.6$ cases.  \label{fig:runII_III_trends}}
\end{figure*}

\begin{figure*}
  \centering
  \includegraphics[width=0.495\linewidth]{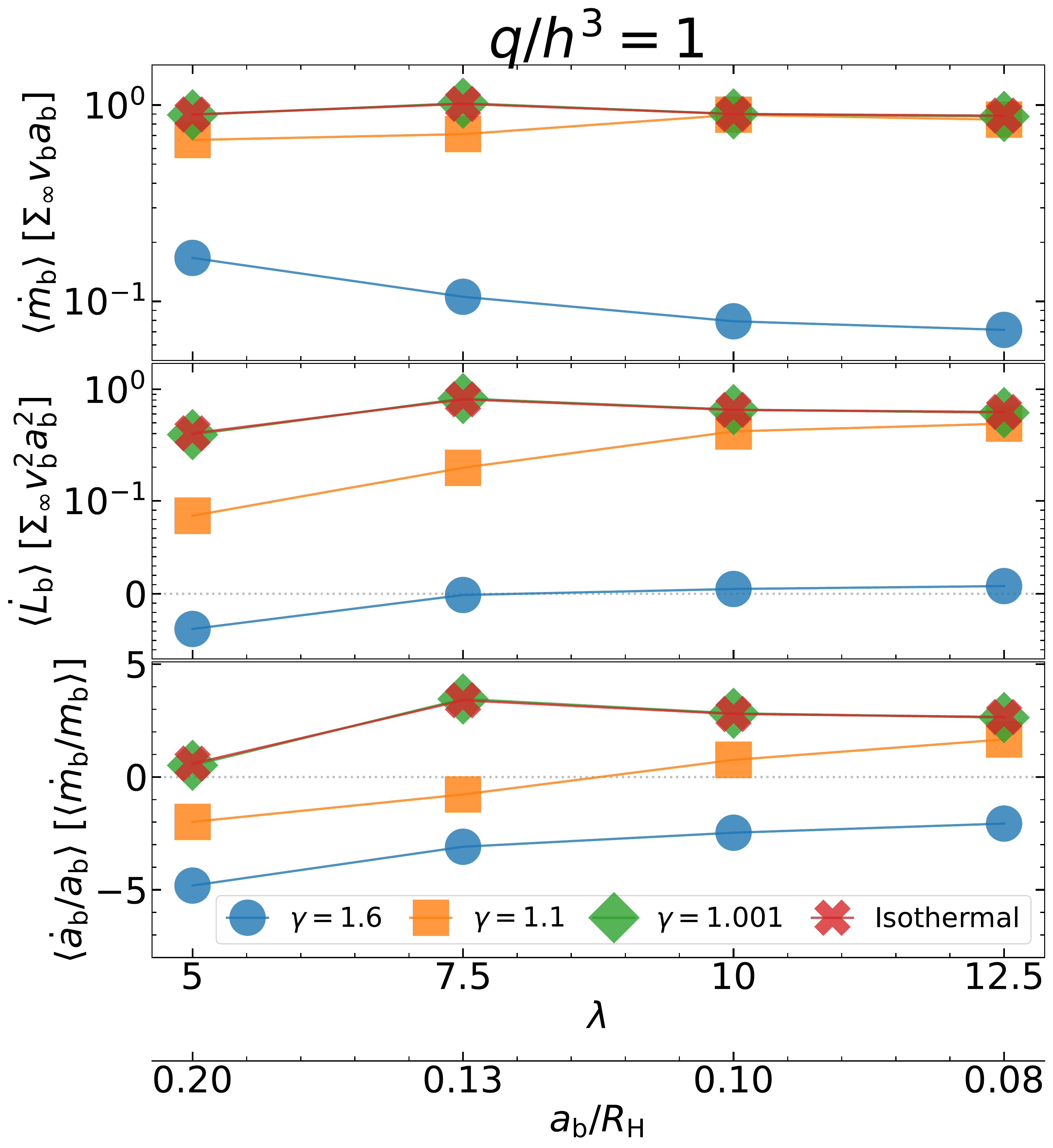}
  \includegraphics[width=0.495\linewidth]{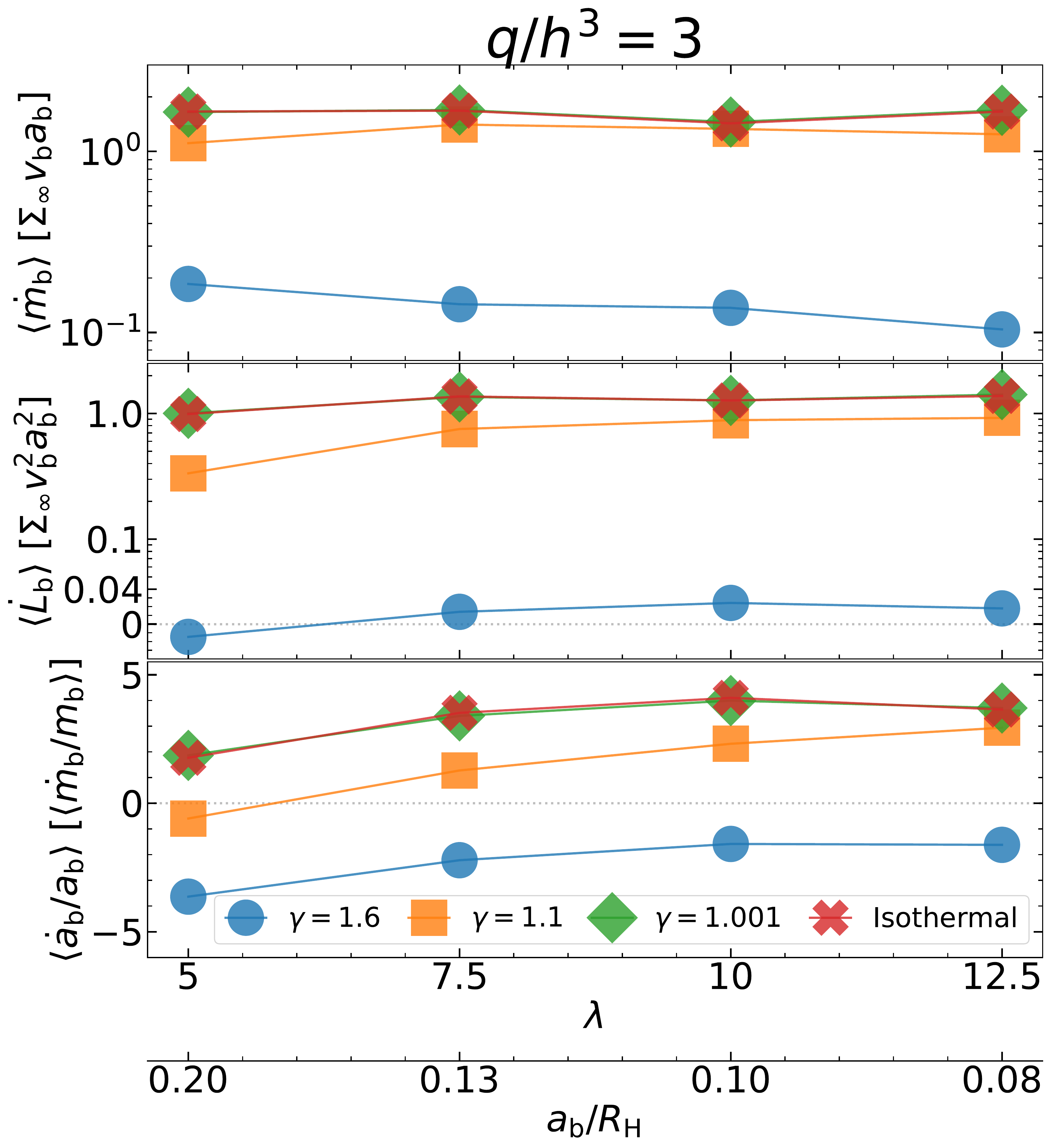}
  \caption{Similar to Fig. \ref{fig:runII_III_trends} but for time-averaged measurements as a function of $\lambda$, for $q/h^3=1$ (\textit{left}) and $q/h^3=3$ (\textit{right}).  The middle panels again combine linear and logarithmic scales along the $y$-axis but the transition point is $10^{-1}$ for $q/h^3=1$ and $0.04$ for $q/h^3=3$.  \label{fig:qth1_3_trends}}
\end{figure*}

\begin{figure*}
  \centering
  \includegraphics[width=0.495\linewidth]{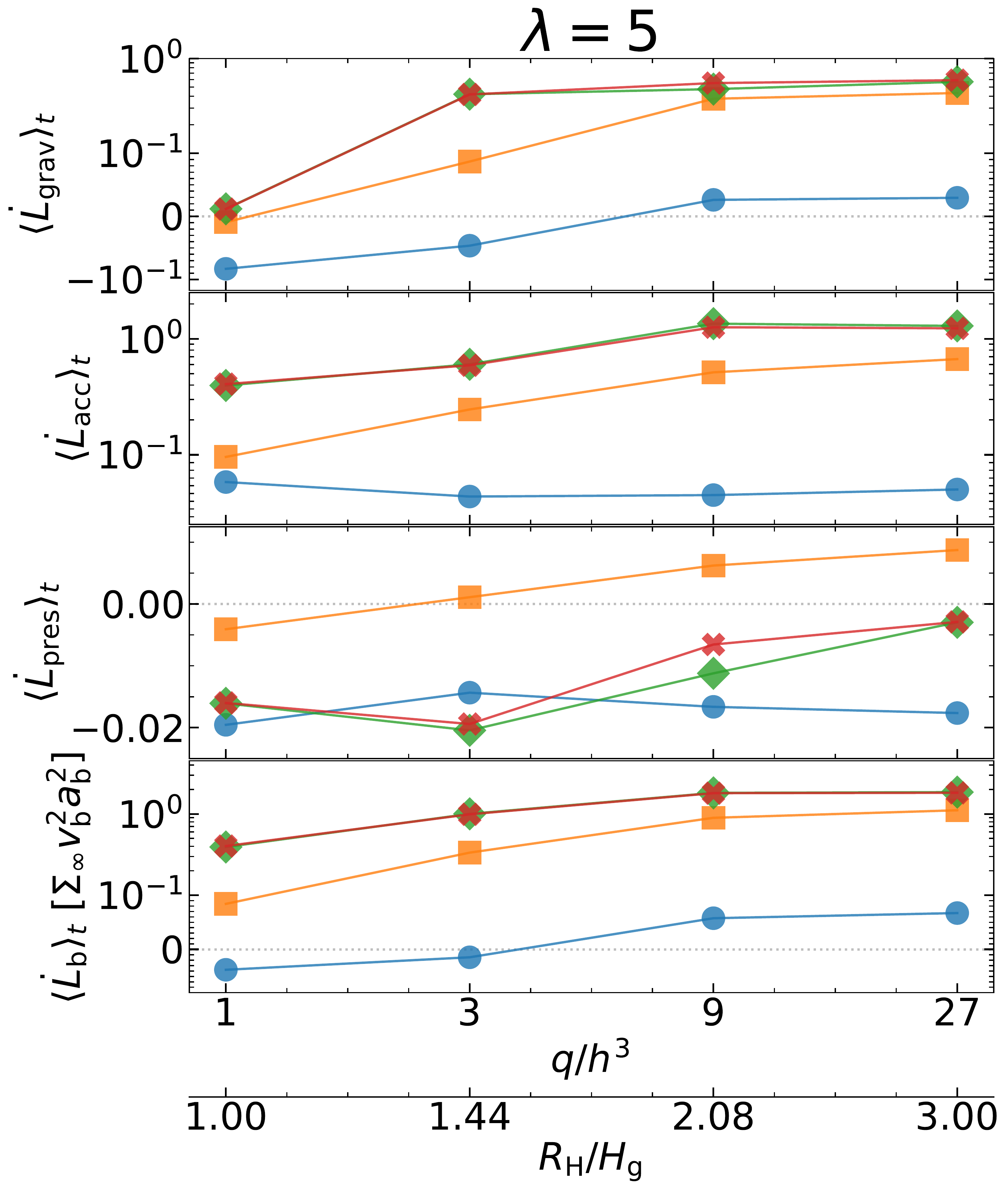}
  \includegraphics[width=0.495\linewidth]{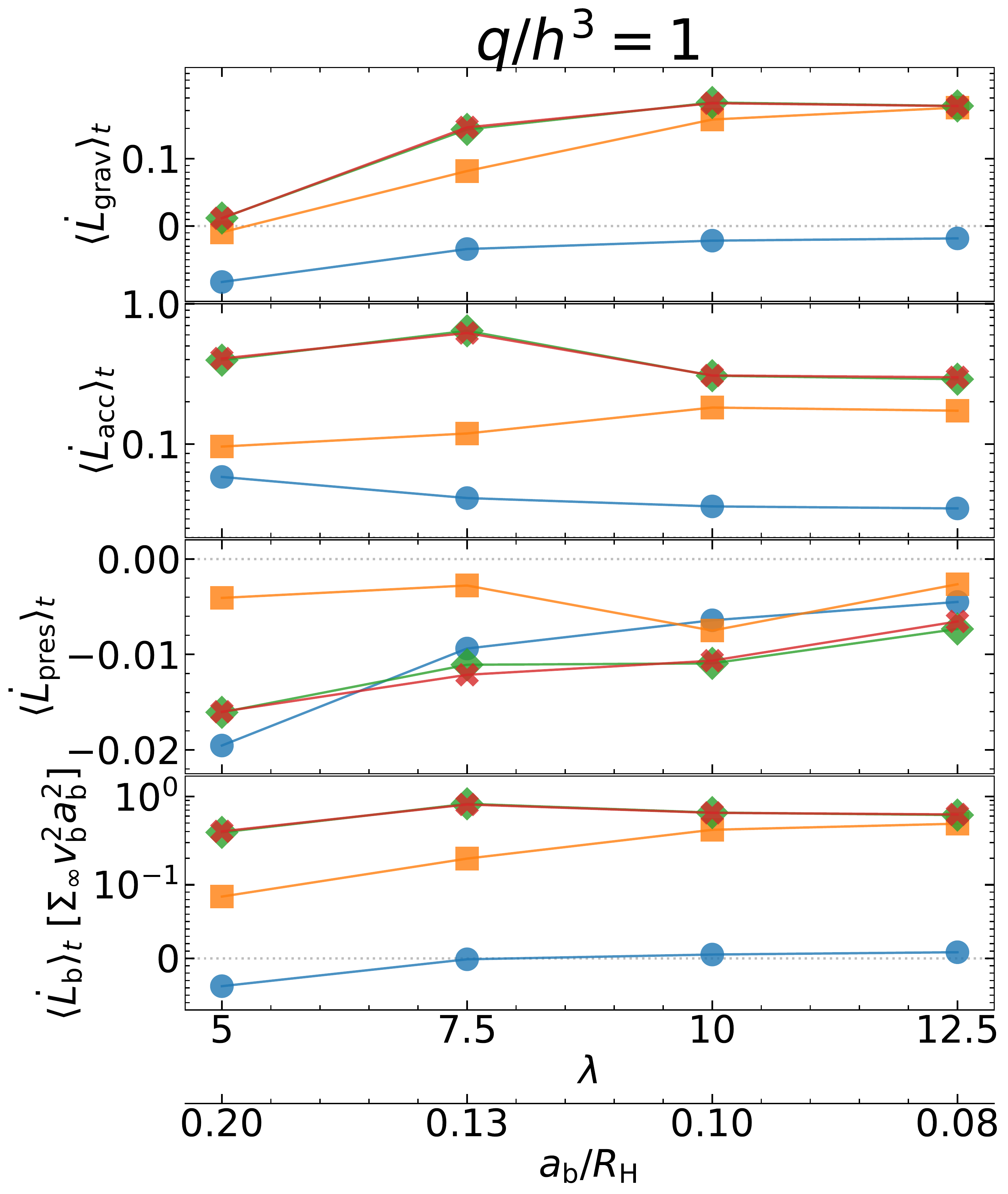}
  \caption{Similar to Figs. \ref{fig:runII_III_trends} and \ref{fig:qth1_3_trends} but showing the time-averaged (from \textit{top} to \textit{bottom}) decomposed and total torques, $\langle\dot{L}_{\rm grav/acc/pres}\rangle$ and $\langle\dot{L}_{\rm b}\rangle$, for \texttt{Run II} series ($\lambda=5$, \textit{left}) and for runs with $q/h^3=1$ (\textit{right}).  All the panels except the third panel combine linear and logarithmic scales along the $y$-axis with a transition point of $10^{-1}$.
  \label{fig:RunII_qth1_trends_dotLs}}
\end{figure*}

\begin{figure*}
  \centering
  \includegraphics[width=0.92\linewidth]{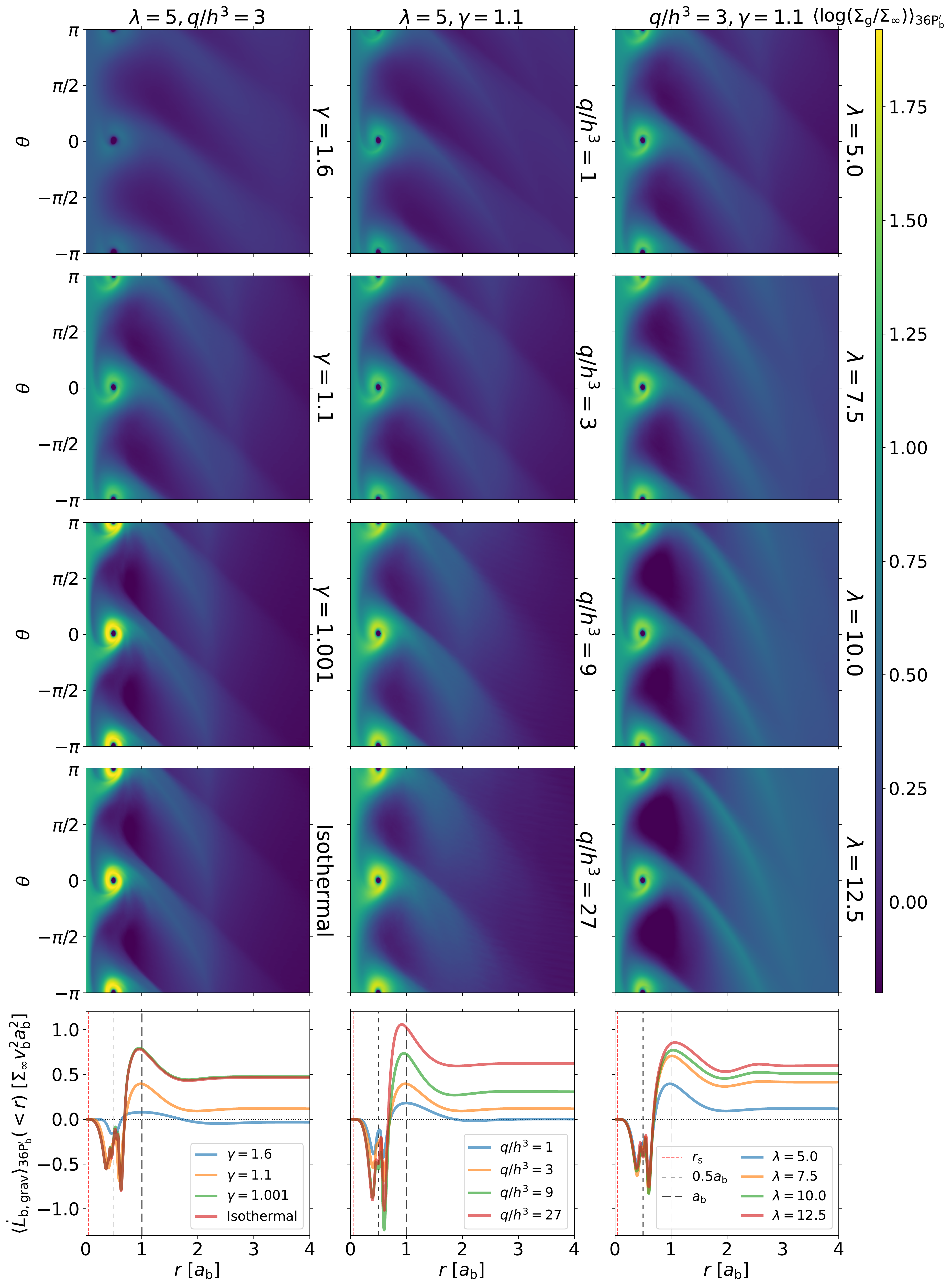}
  \caption{Time-averaged maps of gas density in the polar coordinates (first four rows) and the corresponding cumulative gravitational torque as a function of radius (bottom row) with varying (from \textit{left} to \textit{right}) EOS($\gamma$), $q/h^3$, and $\lambda$.  Several characteristic radial scales are marked by the vertical lines in the bottom row, indicating the sink radius (red short dashed), half binary separation (black medium dashed), and binary separation (black long dashed). \label{fig:CumuTgrav_SgMap}}
\end{figure*}

\begin{figure*}
  \centering
  \includegraphics[width=\linewidth]{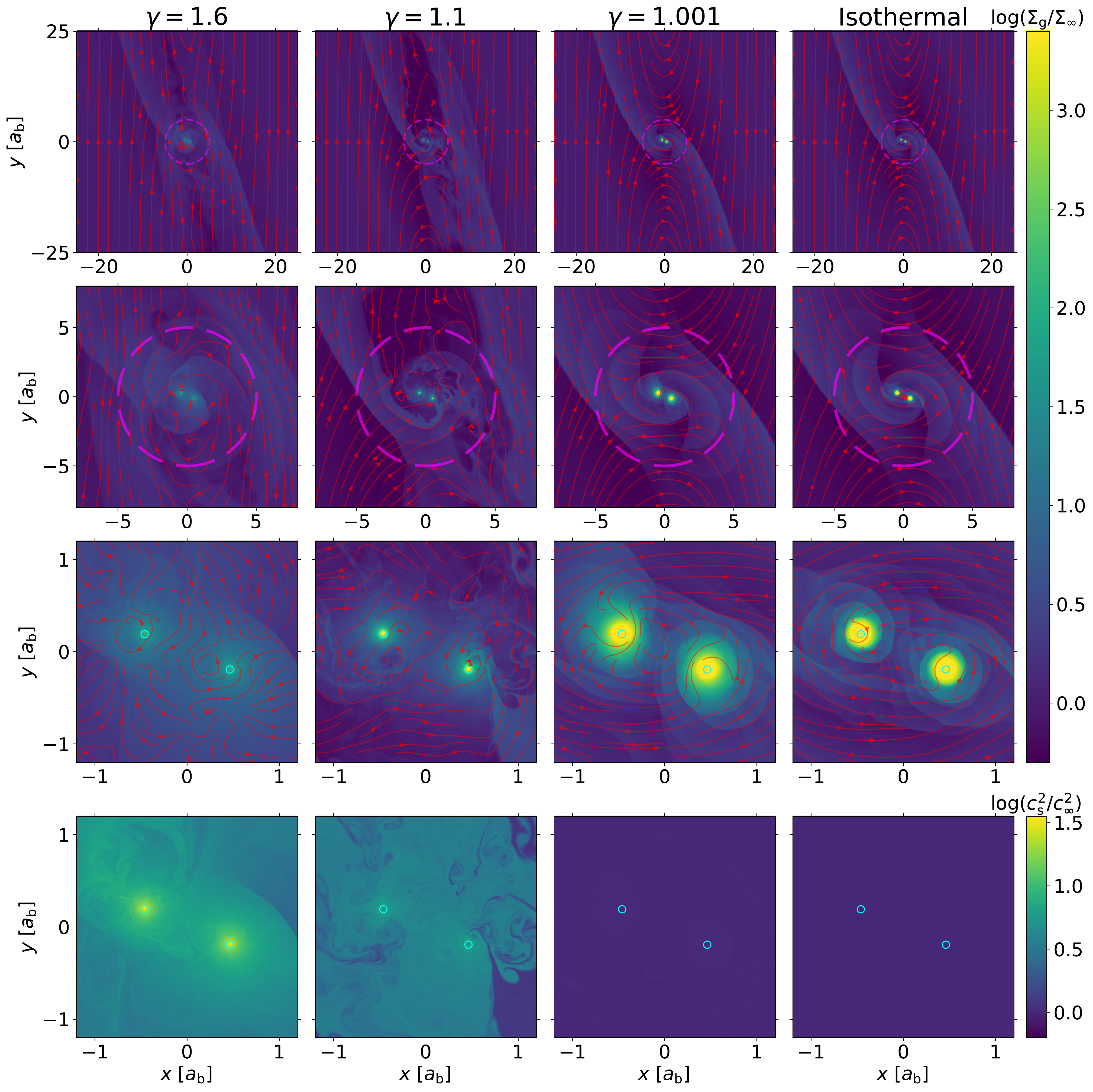}
  \caption{Similar to Fig. \ref{fig:snap_runIII_qth3} but for \texttt{Run II-NA} ($\lambda = 5$) with $q/h^3=1$, where the binary does not accrete gas and has a softening length $\xi_{\rm s}=0.04 a_{\rm b}$, represented by the \textit{cyan solid} circles in the bottom two rows.  \label{fig:snap_runII-NA_qth1}}
\end{figure*}

\begin{figure*}
  \centering
  \includegraphics[width=\linewidth]{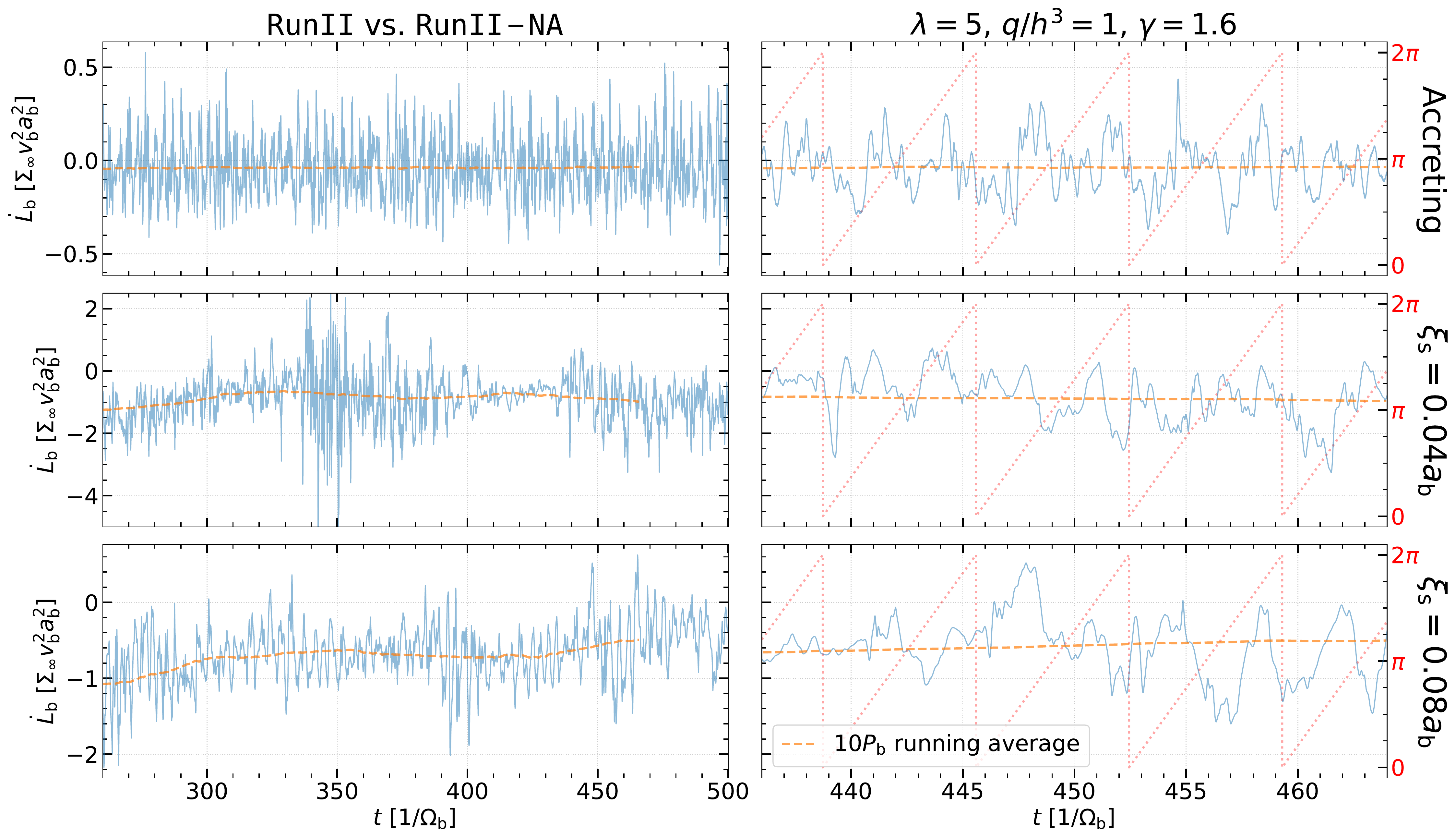}
  \includegraphics[width=\linewidth]{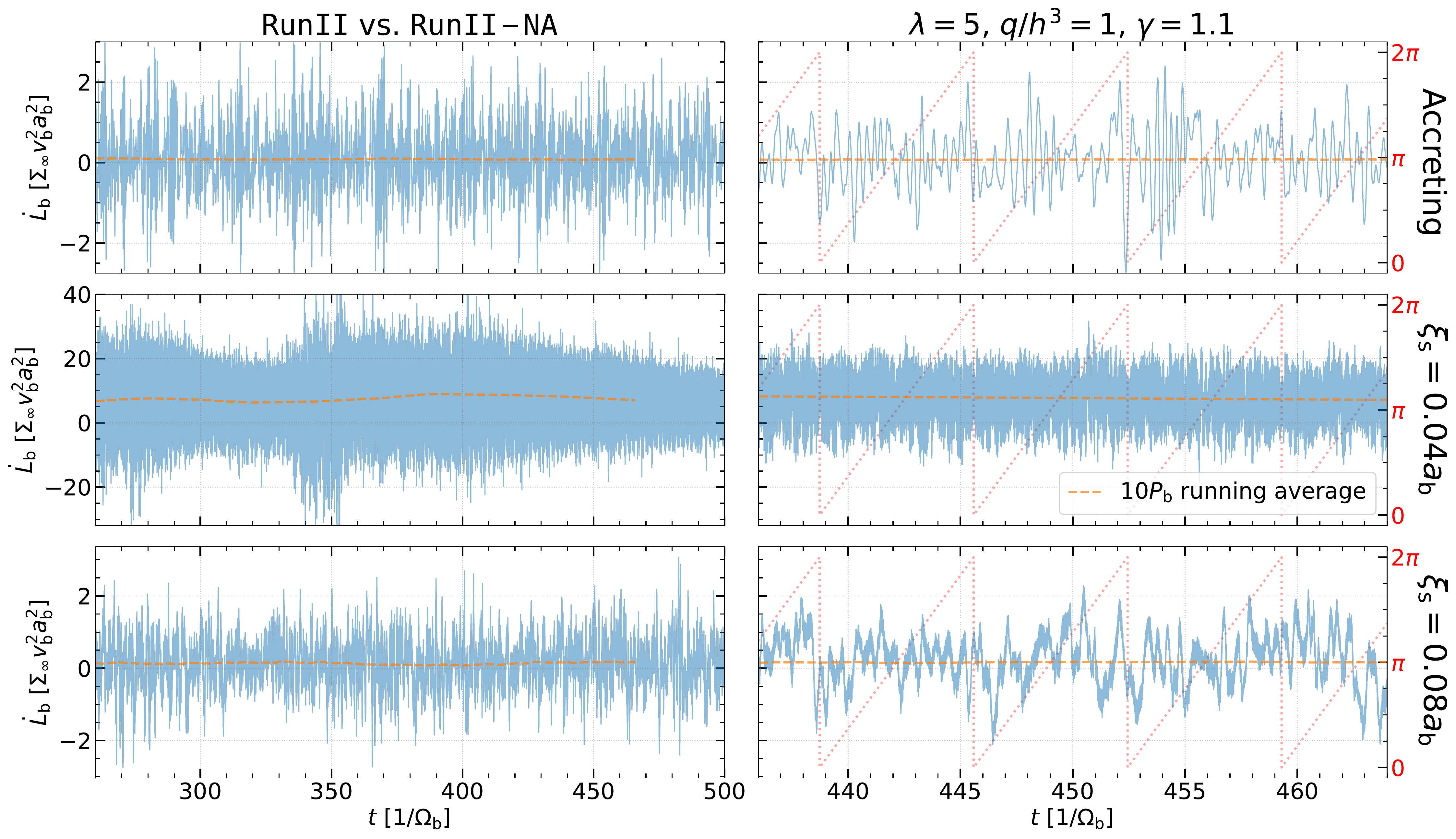}
  \caption{Comparisons of time series of total torques $\dot{L}_{\rm b}$ (\textit{blue solid}) on the binary in \texttt{Run II} (with accretion) and \texttt{Run II-NA} (with no accretion, using gravitational softening of $0.04 a_{\rm b}$ or $0.08 a_{\rm b}$) with $q/h^3=1$ and with $\gamma=1.6$ (\textit{top}) and $\gamma=1.1$ (\textit{bottom}), for the whole time used for time-averaging (\textit{left}) and a small slice of time (\textit{right}).  Each time series is accompanied by the $10P_{\rm b}'$ running average (\textit{dashed orange}) and each slice of time series is accompanied by the binary orbital phase curve with period $P_{\rm b}'$ (\textit{pink dotted}).
  \label{fig:dotL_runII_AccvsNA}}
\end{figure*}

\subsection{Secular Evolution of Binary}
\label{subsec:orbital_evolution}

We employ the same post-processing procedure as described in \citetalias{Li2022} to compute the time series of accretion rate and torques onto the binary in each simulation.  Fig. \ref{fig:runII_III_trends} shows the secular orbital evolution results as a function of $q/h^3$, grouped by the EOS, for our surveys in \texttt{Run II} series ($\lambda=5$) and \texttt{Run III} series ($\lambda=7.5$).  We immediately recognize that all the time-averaged measurements show good agreement between the isothermal cases and the $\gamma=1.001$ cases, which is again reassuring.

We find that the time-averaged accretion rate $\langle\dot{m}_{\rm b}\rangle$, the total torque $\langle\dot{L}_{\rm b}\rangle$, and the binary migration rate $\langle\dot{a}_{\rm b}\rangle/a_{\rm b}$ generally increase with $q/h^3$ and monotonically decrease with $\gamma$, consistent with the findings that the CSDs are more massive and the circumbinary flows are denser in cases with higher $q/h^3$ and/or lower $\gamma$ (see Figs. \ref{fig:snap_runIII_qth3} and \ref{fig:snap_runIII_ga1.1}).  It is remarkable that the accretion rates in the isothermal cases are about one order of magnitude higher than those in the cases with $\gamma=1.6$, whereas $\langle\dot{m}_{\rm b}\rangle$ only increases by a factor of $\sim 3$ when $q/h^3$ increases from $1$ to $27$.  The latter is not surprising since the accretion boost due to the more massive CSDs is attenuated by the much higher binary orbital velocity.  One exception is the high $\langle\dot{m}_{\rm b}\rangle$ in the case of $(\lambda, q/h^3, \gamma)=(7.5, 27, 1.6)$, where the accretion flow is drastically violent.  The CSDs in this case experience severe ram pressure stripping, somewhat similar to the retrograde cases in our \citetalias{Li2022}, leading to intermittent disc-less accretion that boosts the accretion rate irregularly.  As for $\langle\dot{L}_{\rm b}\rangle$ and $\langle\dot{a}_{\rm b}\rangle/a_{\rm b}$, we observe a similar trend that their changes appear to be more prominent with varying EOS than with varying $q/h^3$.

Furthermore, Fig. \ref{fig:qth1_3_trends} shows the binary evolution results as a function of $\lambda$, again grouped by the EOS, for our surveys with $q/h^3=1$ and $3$ across all run series.  
We find that the accretion rate $\langle\dot{m}_{\rm b}\rangle$ slightly decreases with $\lambda$ when $\gamma=1.6$ but varies little with $\lambda$ when $\gamma \leqslant 1.1$.  For smaller $\gamma$, the accretion onto the binary seems to be saturated by the more massive CSDs, whereas for $\gamma=1.6$, $\langle\dot{m}_{\rm b}\rangle$ is buffered by the diffuse CSDs.  Moreover for $\gamma=1.6$, the accretion rate decreases with $\lambda$ largely because the binary orbits at a higher Mach number and intersects with less streamlines as the binary separation becomes smaller relative to $R_{\rm H}$.

On the other hand, we find that both the total torque $\langle\dot{L}_{\rm b}\rangle$ and the binary migration rate $\langle\dot{a}_{\rm b}\rangle/a_{\rm b}$ generally slightly increases with $\lambda$.  In addition, all these measurements tend to flatten out at $\lambda \gtrsim 10$.  This trend is understandable since a large region within $R_{\rm H}$ feels the binary as a single object as $\lambda$ increases, and the hydrodynamical evolution eventually reduces to that around a single object when $\lambda \to \infty$.

Regarding the direction of binary migration, i.e., the sign of $\langle\dot{a}_{\rm b}\rangle$, we find that all runs with $\gamma=1.6$ produce contracting binaries, consistent with the findings in our \citetalias{Li2022}.  All runs with $\gamma=1.001$ and with the isothermal EOS produce expanding binaries, consistent with previous isothermal studies \citep[e.g.,][]{LiYaPing2021,LiYaPing2022}.  That said, the trends identified in our results indicate that wider binaries ($\lambda < 5$) with the isothermal EOS may be instead contracting, roughly consistent with recent work that focuses on the binary separation \citep{Dempsey2022}.  The cases with $\gamma=1.1$ produce inspiral binaries at smaller $q/h^3$ and/or $\lambda$, and outspiral binaries at higher $q/h^3$ and/or $\lambda$.  We further find that the total angular momentum transferred to the binary is negative for $\gamma=1.6$ and low $q/h^3$ and/or $\lambda$.  Otherwise, the binary is always fed by positive angular momentum.  All these findings indicate that the cases where the binary is able to sustain more massive CSDs yield more positive torques and thus less negative $\langle\dot{a}_{\rm b}\rangle/a_{\rm b}$. 

\section{Torques}
\label{sec:torques}

In this section, we discuss how different torque components contribute to the total torque during the secular binary orbital evolution.  The gravitational torque $\langle\dot{L}_{\rm b,grav}\rangle$ is often perceived as the largest and most important torque.  Section \ref{subsec:torque_decomposition} demystifies such a perception, followed by a detailed comparison of the spatial distribution of $\langle\dot{L}_{\rm b,grav}\rangle$ in Section \ref{subsec:Tgrav_map}.

\subsection{Torque Decomposition}
\label{subsec:torque_decomposition}

Fig. \ref{fig:RunII_qth1_trends_dotLs} shows the decomposed torque contributions from gravity (i.e., dynamical friction), $\langle\dot{L}_{\rm b,grav}\rangle$, and from the hydrodynamical forces associated with gas accretion and pressure, $\langle\dot{L}_{\rm b,acc}\rangle$ and $\langle\dot{L}_{\rm b,pres}\rangle$, for the cases with $\lambda=5$ and $q/h^3=1$.

We find that $\langle\dot{L}_{\rm b,acc}\rangle$ is always positive and its magnitude is often comparable to that of $\langle\dot{L}_{\rm b,grav}\rangle$.  In particular, in the cases with high accretion rates, (i.e., larger $q/h^3$, higher $\lambda$, and/or smaller $\gamma$), both $\langle\dot{L}_{\rm b,acc}\rangle$ and $\langle\dot{L}_{\rm b,grav}\rangle$ are positive and the former is sometimes larger than the latter.  This finding indicates that the torque due to the hydrodynamical force from accretion is as essential as the gravitational torque in determining the long-term binary orbital evolution.  Neglecting $\langle\dot{L}_{\rm b,acc}\rangle$ may underestimate the total torque exerted on the binary and in some cases lead to a different binary migration direction (i.e., softening to hardening).

Unlike $\langle\dot{L}_{\rm b,grav}\rangle$ and $\langle\dot{L}_{\rm b,acc}\rangle$ that monotonically decrease with $\gamma$, the torque due to the hydrodynamical force from pressure $\langle\dot{L}_{\rm b,pres}\rangle$ shows a more complex $\gamma$-dependency.  Such a feature is likely due to the $\gamma$-dependency of the spiral morphology, i.e., the smaller the $\gamma$, the tighter the small spiral shocks are winded in the CSDs.  In most cases (except some with $\gamma=1.1$), $\langle\dot{L}_{\rm b,pres}\rangle$ is negative.  The magnitude of $\langle\dot{L}_{\rm b,pres}\rangle$ is typically small but can be significant when $\gamma=1.6$, where the other two torque components are comparably small, as noted in our \citetalias{Li2022}.

\subsection{Spatial Distribution of Gravitational Torque}
\label{subsec:Tgrav_map}

In this section, we follow our \citetalias{Li2022} and construct the time-averaged maps (over the last $36 P_{\rm b}'$; 4 snapshots per $\Omega_{\rm b}^{-1}$) for the gas surface density and the gravitational torque surface density in the vicinity of the binary for all the cases shown in Figs. \ref{fig:snap_runIII_qth3}-\ref{fig:snap_qth3_ga1.1}.  Fig. \ref{fig:CumuTgrav_SgMap} shows the $\langle \Sigma_{\rm g} \rangle$ maps in the polar coordinates and compares the corresponding cumulative gravitational torque as a function of radius.

The $\langle \Sigma_{\rm g} \rangle$ maps reveal the persistent non-axisymmetric structures that affect the secular binary orbital evolution, including the CSDs and the large spirals shocks extending to a few $a_{\rm b}$.  Any persistent density enhancements within the phase range ($0$, $\pi/2$) and ($-\pi$, $-\pi/2$) are ``leading'' with respect to the binary and contribute positive $\langle\dot{L}_{\rm b,grav}\rangle$, whereas those in the phase range ($\pi/2$, $\pi$) and ($-\pi/2$, $-\pi$) are ``trailing'' and contribute negative $\langle\dot{L}_{\rm b,grav}\rangle$.  Generally speaking, we find that the part of the CSDs inside the binary orbit ($0.2 a_{\rm b} \lesssim r < 0.5 a_{\rm b}$) and the inner part of the large spiral shocks ($a_{\rm b} \lesssim r \lesssim 2 a_{\rm b}$) provide negative torques, while the remainders of the CSDs ($0.5 a_{\rm b} < r \lesssim 0.8 a_{\rm b}$) and the large spiral shocks ($\gtrsim 2 a_{\rm b}$) provide positive torques.  Also, the net torque of the CSDs is positive and the net torque of the large spiral shocks is negative.

We find that the total gravitational torque $\langle\dot{L}_{\rm b,grav}\rangle$ decreases with $\gamma$ and increases with $q/h^3$.  When $\gamma$ is smaller or $q/h^3$ is larger, both the CSDs and the large spiral shocks are more massive/denser, but the positive net torque from the CSDs dominates the total torque.  Only in most cases with $\gamma=1.6$ or a few cases with $q/h^3=1$, does the negative net torque from the large spiral shocks dominates the total torque.  In addition, Fig. \ref{fig:CumuTgrav_SgMap} shows that the radial profile of the cumulative gravitational torque with $\gamma=1.001$ overlaps well with that of the isothermal case, which reassures us not only the robustness of the code but also our post-processing analyses.

We further find that $\langle\dot{L}_{\rm b,grav}\rangle$ increases with $\lambda$.  This increase largely occurs in the region from $0.5 a_{\rm b}$ to $\sim a_{\rm b}$ since the cumulative torque profile inside $0.5 a_{\rm b}$ or outside $\sim a_{\rm b}$ changes little when $\lambda$ varies.  As $\lambda$ becomes larger, the CSDs are only slightly more massive but the voids within $a_{\rm b}$ are emptier (see also Section \ref{subsec:flow_field}) and clear out more trailing regions, resulting in a more positive net torque.  Even for $\lambda \gtrsim 10$, we find that the persistent non-axisymmetric structures only extends out to a few $a_{\rm b}$ (i.e., $\sim 3 a_{\rm b}$, a factor of a few smaller than $R_{\rm H}$), where the cumulative torque profile already becomes flat and roughly equals to the total gravitational torque.

\begingroup 
\setlength{\medmuskip}{0mu} 
\setlength\tabcolsep{4pt} 
\setcellgapes{3pt} 
\begin{table}
  \nomakegapedcells
  \caption{Results for Simulations with Non-accreting Binaries\\ (\texttt{Run II-NA}, $\lambda=5$, $q/h^3=1$)} \label{tab:runs-NA}
  \makegapedcells 
  \linespread{1.025} 
  \begin{tabular}{ll|rr}
    \hline
    $\gamma$
    & \makecell[c]{$\xi_{\rm s}$}
    & \makecell[c]{$\langle\dot{L}_{\rm b}\rangle$}
    & \makecell[c]{$\displaystyle \frac{\langle\dot{a}_{\rm b}\rangle}{a_{\rm b}}$}
    \\
    
    & \makecell[c]{[$a_{\rm b}$]}
    & \makecell[c]{\footnotesize $\displaystyle \left[ \Sigma_\infty v_{\rm b}^2 a_{\rm b}^2 \right]$}
    & \makecell[c]{\small $\displaystyle \left[ \frac{\Sigma_\infty a_{\rm b} v_{\rm b}}{m_{\rm b}}  \right]$}
    \\
    (1)
    & \makecell[c]{(2)}
    & \makecell[c]{(3)}
    & \makecell[c]{(4)}
    \\
    \hline\hline
        & $10^{-8}$ & $-0.04$ & $-0.80$ \\ \cline{2-4}
    1.6 & $0.04$ & $-0.89$ & $-7.15$ \\
        & $0.08$ & $-0.67$ & $-5.37$ 
    \\
    \Xhline{3\arrayrulewidth}
        & $10^{-8}$ & $0.08$ & $-1.32$ \\ \cline{2-4}
    1.1 & $0.04$ & $7.50$ & $59.97$ \\
        & $0.08$ & $0.13$ & $1.08$ 
    \\
    \Xhline{3\arrayrulewidth}
          & $10^{-8}$ & $0.39$ & $0.46$ \\ \cline{2-4}
    1.001 & $0.04$ & $0.14$ & $1.10$ \\ 
          & $0.08$ & $0.15$ & $1.22$
    \\
    \Xhline{3\arrayrulewidth}
      & $10^{-8}$& $0.40$& $0.53$ \\ \cline{2-4}
    1 & $0.04$ & $-0.004$ & $-0.03$ \\
      & $0.08$ & $-0.0002$ & $-0.002$
    \\
    \hline
  \end{tabular} \\
  \begin{flushleft}
    {\large N}OTE ---Columns: 
    (1) $\gamma$ in the $\gamma$-law EOS (the cases with $\gamma=1$ are isothermal runs);  
    (2) gravitational softening length (the rows with $\xi_{\rm s} = 10^{-8} a_{\rm b}$ are \textit{reference} rows from the standard \texttt{Run-II} with accreting binaries); 
    (3) time-averaged rate of change of the binary angular momentum;
    (4) binary semimajor axis change rate.
    \\
  \end{flushleft}
\end{table}
\endgroup

\section{Comparisons with Non-accreting Black Holes}
\label{sec:non-accreting}

In this section, we perform experiments on the evolution of embedded BBHs where the binary does not accrete at all.  Non-accreting binaries are of interest because certain physical processes (e.g., BH feedback) may significantly hamper or even halt the accretion.  We thus employ idealized non-accreting binaries as a first attempt to study the effects of these processes, which are challenging to model directly.

Table \ref{tab:runs-NA} summarizes the key parameters and results of our experiments (the values are time-averaged over the same time period as specified in Section \ref{subsec:setups}).  For each case in \texttt{Run II} ($\lambda=5$) with $q/h^3=1$, we conduct two simulations with non-accreting binaries with a softening length of $\xi_{\rm s} = 0.04 a_{\rm b}$ and $\xi_{\rm s} = 0.08 a_{\rm b}$, respectively.  There is no sink cell in these experiments and all cells in the computational domain contribute to the gravitational torque on the binary via force
\begin{equation}
  \bm{f}_{\rm grav,i} = - G m_k \frac{\bm{r}_i - \bm{r}_k}{\left[ (\bm{r}_i - \bm{r}_k)^2 + \xi_{\rm s}^2 \right]^{1.5}},
\end{equation}
where $i=1$ or $2$ labels a binary component, $m_k = \Sigma_{\rm g} \delta^2$ is the gas mass in the $k$-th cell, and $\delta$ is the cell size of the finest level available at $\bm{r}_k$.

Figure \ref{fig:snap_runII-NA_qth1} shows the flow structure in \texttt{Run II-NA} with $\xi_{\rm s} = 0.04 a_{\rm b}$.  Since there is no sink cells, gas simply accumulates in the potential wells as blobs around the binary components, does not form consistent CSDs, and appears to be largely pressure supported.  The maximum gas density around the binary components is orders of magnitude higher than that in the counterpart simulations with sink cells.  Also, we find a similar dependence on $\gamma$ where the gas blobs become more massive as $\gamma$ decreases.

The circumbinary flow remains prograde in the isothermal case and in the case with $\gamma=1.001$, but the gas may be deflected when getting too close to the dense blobs.  For $\gamma \geqslant 1.1$, the flow around the binary is stochastic since the gas is heated up but no accretion is available to exhaust the excessive energy.  In all cases, even the flow outside $R_{\rm H}$ differs from their counterparts with accreting binaries.  In particular, the horseshoe flows are heavily disrupted when $\gamma=1.6$.  For the cases with $\xi_{\rm s} = 0.08 a_{\rm b}$, the flow structure is broadly similar, with smaller but slightly less dense blobs.

The secular orbital evolution of non-accreting binaries depends only on the gravitational torque:
\begin{equation}
  \frac{\langle \dot{a}_{\rm b} \rangle}{a_{\rm b}} = 2 \frac{\langle \dot{L}_{\rm b} \rangle}{L_{\rm b}} \ \ \ \text{and}\ \ \
  \dot{L}_{\rm b} = \dot{L}_{\rm grav} = \mu_{\rm b} \bm{r}_{\rm b} \times (\bm{f}_{\rm grav,1} - \bm{f}_{\rm grav,2}).
\end{equation}
The gravity $f_{\rm grav,i}$ is largely determined by the dense gas blobs, making $\bm{f}_{\rm grav,1} - \bm{f}_{\rm grav,2}$ almost parallel to $\bm{r}_{\rm b}$.  Thus, the value of $\dot{L}_{\rm grav}$ is sensitive to the deviation of the gas density structure from symmetry, which can easily be stochastic.

The resulting torque on non-accreting binaries is intrinsically stochastic regardless of timescale.  Fig. \ref{fig:dotL_runII_AccvsNA} compares the time series of $\dot{L}_{\rm b}$ in \texttt{Run II} and \texttt{Run II-NA} with $\gamma=1.6$ and $\gamma=1.1$.  Although the running averages of $\dot{L}_{\rm b}$ in \texttt{Run II} are nearly constant and are in good agreement with the time-averaged results in Table \ref{tab:runs-NA} (as also seen in our \citetalias{Li2022}), the running averages in \texttt{Run II-NA} show non-negligible irregular variations over hundreds of $\Omega_{\rm b}^{-1}$.  Moreover, the instantaneous variation amplitudes of $\dot{L}_{\rm b}$ are very sensitive to the softening length and are much larger in the cases of $\xi_{\rm s}=0.04 a_{\rm b}$, where the gas blobs are denser and more concentrated.

When $\gamma$ is lower (e.g., $\lesssim 1.1$), the maximum gas surface density in the blobs around non-accreting binaries exceeds $10^4 \Sigma_{\infty}$ (as a comparison, max$(\Sigma_{\rm g}) \sim 10^2 \Sigma_{\infty}$ when $\gamma=1.6$).  Therefore, even slight asymmetries originated from such dense blobs lead to considerable torques, which vary stochastically with frequencies much higher than the binary orbital frequency (see Fig. \ref{fig:dotL_runII_AccvsNA}).  Also, the relative power on these high frequencies again depends on the softening length (i.e., less power when $\xi_{\rm s}$ is larger since the blobs are less concentrated).

The secular orbital evolution of non-accreting binaries thus significantly differs from that of accreting binaries (see also Table \ref{tab:runs-NA}) and does not present a monotonic trend as a function of $\gamma$.  Additionally, the secular evolution is affected by the choice of softening length $\xi_{\rm s}$, especially in the cases where the dense gas blobs are highly concentrated within the softening length (e.g., $\gamma=1.1$).  That said, for (nearly-)isothermal cases (including $\gamma=1.001$), the gas remains cool such that the dense gas blobs extend far beyond $\xi_{\rm s}$, making the secular evolution less dependent on the softening length.

\section{Summary}
\label{sec:summary}

In this follow-up study to our \citetalias{Li2022}, we perform a suite of 2D simulations to study the evolution of prograde equal-mass circular binary black holes (BBHs) embedded in AGN discs with an extensive coverage of parameter space.  As discussed in Section \ref{subsec:schemes}, the flow dynamics and binary evolution depend on three dimensionless parameters: $\gamma$ (in the $\gamma$-law EOS), $q/h^3\equiv (m_{\rm b}/M)(R/H_{\rm g})^3 \equiv (R_{\rm H}/H_{\rm g})^3$ and $\lambda = R_{\rm H}/a_{\rm b}$.  As in \citetalias{Li2022}, we focus on accreting binaries, resolving the flow down to $\lesssim 1\%$ of the Hill radius around each binary component (with finest cells $\lesssim 0.1\%$ of Hill radius; Sections \ref{sec:methods} to \ref{sec:torques}), although we also consider non-accreting binaries (Section \ref{sec:non-accreting}) in this paper.  Our key findings are as follows:

\begin{enumerate}
  \item The accretion flow, particularly the morphology and the thermodynamics of the CSDs, considerably depends on $\gamma$, $q/h^3$, and $\lambda$ (see Figs. \ref{fig:snap_runIII_qth3}, \ref{fig:snap_runIII_ga1.1}, and \ref{fig:snap_qth3_ga1.1}).  The CSDs are generally more massive with decreasing $\gamma$ and with increasing $q/h^3$ and $\lambda$.  Also, they are generally hotter and more turbulent with increasing $\gamma$, $q/h^3$, and $\lambda$, though the $\lambda$-dependency is somewhat weaker.  Moreover, the time-averaged accretion rate $\langle \dot{m}_{\rm b} \rangle$ (in units of $\Sigma_{\infty} a_{\rm b} v_{\rm b}$) monotonically decreases with $\gamma$ and increases with $q/h^3$. 
  \item For all the accreting binary cases considered in this paper, we find that the binary contracts when $\gamma=1.6$ and expands when $\gamma=1.001$ or when using the isothermal EOS (see Figs. \ref{fig:runII_III_trends} and \ref{fig:qth1_3_trends} and Table \ref{tab:runs}).  The time-averaged binary migration rate $\langle \dot{a}_{\rm b} \rangle/a_{\rm b}$ generally decreases with $\gamma$ and increases with $q/h^3$ and $\lambda$.  The typical magnitude of $\langle \dot{a}_{\rm b} \rangle/a_{\rm b}$ is of the order of a few times mass doubling rate $\langle \dot{m}_{\rm b} \rangle/m_{\rm b}$.
  \item When accretion is allowed, the torque associated with accretion $\langle\dot{L}_{\rm b,acc}\rangle$ often has a magnitude comparable to the gravitational torque $\langle\dot{L}_{\rm b,grav}\rangle$ and is thus essential in determining the binary orbital evolution (see Fig. \ref{fig:RunII_qth1_trends_dotLs}).  When the CSDs are diffuse and $\langle \dot{m}_{\rm b} \rangle$ is moderate (i.e., higher $\gamma$ or smaller $q/h^3$), the torque associated with pressure $\langle\dot{L}_{\rm b,pres}\rangle$ can also have a comparable contribution to the total torque on the binary.
  \item For prograde equal-mass circular binaries, the gravitational torque $\langle\dot{L}_{\rm b,grav}\rangle$ is largely determined by the persistent non-axisymmetric flow structures within a few $a_{\rm b}$, regardless of $\lambda$ (see Fig. \ref{fig:CumuTgrav_SgMap}).  Similar to the parameter dependence of the accretion flow, we find that $\langle\dot{L}_{\rm b,grav}\rangle$ decreases with $\gamma$ and increases with $q/h^3$ and with $\lambda$.
  \item For non-accreting binaries (see Section \ref{sec:non-accreting}), the flow structure substantially differs from that around accreting binaries, where a pressure-supported dense blob forms around each binary component instead of a CSD (see Fig. \ref{fig:snap_runII-NA_qth1}).  The consequent torque on the binary and its orbital evolution vary stochastically, distinct from those seen for the accreting binaries, and are dependent on the gravitational softening length $\xi_{\rm s}$ (see Fig. \ref{fig:dotL_runII_AccvsNA} and Table \ref{tab:runs-NA}).  The torque/evolution is most sensitive to $\xi_{\rm s}$ at moderate $\gamma$ where the gas blobs are cool enough to be dense but are also hot enough to be highly concentrated within the softening length.
\end{enumerate}

The survey in this work provides a complement to the parameter space covered by our prior work \citepalias{Li2022}, which focused on $\gamma=1.6$, but covers a range of binary mass ratios and eccentricities.  These findings are qualitatively consistent with the studies from other groups when they overlap \citep{LiYaPing2021, LiYaPing2022, Dempsey2022}.  Since the numerical scheme employed in this follow-up paper remains the same as \citetalias{Li2022}, our results are still subject to limitations from the 2D local shearing box approximation and from missing physics like viscosity, magnetic field, and so forth.  That said, these 2D simulations allow us to explore a large parameter space and evolve the binary for thousands of $\Omega_{\rm b}^{-1}$ so that we can quantify their secular orbital evolution.  Future 3D studies with more realistic physical ingredients, both local and global, are required to better understand the interactions between BBHs and AGN discs.

\section*{Acknowledgements}
This work has been supported in part by the NSF grant AST-2107796 and the NASA grant 80NSSC19K0444.  Resources supporting this work were provided by the NASA High-End Computing (HEC) Program through the NASA Center for Climate Simulation (NCCS) at Goddard Space Flight Center.  

RL thanks Kaitlin Kratter, Hui Li, Diego {Mu{\~n}oz}, Ya-Ping Li, Adam Dempsey, Zoltan Haiman, Yan-Fei Jiang, Paul Duffell, and Barry McKernan for inspiring discussions and useful conversations.  This research was supported in part by the National Science Foundation under Grant No. NSF PHY-1748958.

\section*{Data Availability}
The simulation data underlying this paper will be shared on reasonable request to the corresponding author.

\bibliographystyle{mnras}
\bibliography{refs}




\bsp    
\label{lastpage}
\end{document}